\documentclass[prd,aps,showpacs,nofootinbib,floats,floatfix,twocolumn,letterpaper]{revtex4}

\usepackage{amssymb,graphicx}
\usepackage{epsfig}
\usepackage[usenames]{color}

\def\rmd{{\rm d}}

\begin{document}

\title{Post-merger electromagnetic emissions from disks perturbed by binary black holes}

\author{Matthew Anderson${}^1$, Luis Lehner${}^{2,3,4}$, 
Miguel Megevand${}^1$ and David Neilsen${}^5$}

\affiliation{
${}^1$Department of Physics and Astronomy, Louisiana State University, Baton Rouge, LA 70803-4001 \\
${}^2$Perimeter Institute for Theoretical Physics,Waterloo, Ontario N2L 2Y5, Canada\\
${}^3$Department of Physics, University of Guelph, Guelph, Ontario N1G 2W1, Canada\\
${}^4$CIFAR, Cosmology \& Gravity Program\\
${}^5$Department of Physics and Astronomy, Brigham Young University, Provo, UT 84602 \\
}

\date{\today}

%
%
\begin{abstract}
We simulate the possible emission from a disk perturbed by a
recoiling super-massive black hole.  To this end, we study radiation transfer
from the system incorporating bremsstrahlung emission from a Maxwellian plasma
and absorption given by Kramer's opacity law modified to incorporate blackbody
effects.  We employ this model in the radiation transfer integration to
compute the luminosity at several frequencies, and compare with previous
bremsstrahlung luminosity estimations from a transparent limit (in which the 
emissivity is integrated over the computational domain and over all frequencies)
and with a simple thermal emission model.  We find close agreement between the
radiation transfer results and the estimated bremsstrahlung luminosity from
previous work for electromagnetic signals above $10^{14}$~Hz.  For lower
frequencies, we find a self-eclipsing behavior in the disk, resulting in a
strong intensity variability connected to the orbital period of the disk.
\end{abstract}

\maketitle

%
%
\section{Introduction:}
The near possibility of detecting gravitational waves to study
astrophysical systems has spurred efforts to understand such systems
theoretically to aid in the detection and interpretation of obtained
signals (see, for instance,
\cite{Aylott:2009tn,Aylott:2009ya,Shoemaker:2008pe,Baumgarte:2006en,
Pan:2007nw,Ajith:2007kx}
and references cited therein for further details and examples).
Systems
that radiate strongly in both electromagnetic and gravitational wave
bands are particularly interesting since the combined information
will provide unprecedented access to a number of rich phenomena
(e.g.~\cite{Haiman:2008zy,Bloom:2009vx,Blecha:2008mg}.)  
Electromagnetic radiation from compact objects can be (partially or
completely) absorbed and scattered by dust and other surrounding
material.  Gravitational radiation, on the other hand, is generated
in the central engine, and these waves do
not interact with dust or material in the surrounding environment.

One such system involves the collision of a binary black hole within
a circumbinary disk
\cite{2002ApJ...567L...9A,2003MNRAS.340..411L,Milosavljevic:2004cg}.  
For a generic black hole binary, the gravitational waves are radiated
asymmetrically.  As these waves carry energy and momentum, the resultant
black hole can have a kick or recoil 
velocity~\cite{kick1,kick3,kick5,kick7,kick9,kick10}, which in turn will
induce shocks in the disk. 
A second source of perturbations in the disk comes from the loss of 
gravitational mass with the emission of gravitational radiation, weakening
the gravitational potential felt by the disk.  These effects, as
discussed in~\cite{kickdisk3,kickdisk1,kickdisk2}, can induce
possible emissions and some of these have been investigated in part
in~\cite{kickdisk1,kickdisk2,massloss,kickdisk,Corrales:2009nv}.  These
works study the dynamics of the disks when either mass reduction
(due to the energy lost by the system) 
or recoil velocity 
(due to the asymmetric flux of gravitational radiation) 
effects are included and
present preliminary estimates of  emissions from the system
with the goal of understanding possible observations in the
electromagnetic spectrum.

In the present work, we further investigate electromagnetic radiation 
from shocked disks by directly computing the electromagnetic 
emission using radiation transfer.  We build on our earlier work on
disks~\cite{kickdisk}, hereafter referred to as Paper~I, 
where we followed the evolution of
disks when the central black hole was given different
recoil velocities. We use this data for the evolution of  disks
to calculate the electromagnetic spectrum using radiation transfer.
In this method, we
assume simple models for the emission and absorption of electromagnetic
radiation and solve for the intensities by integrating the radiation
transfer equation along geodesics of the spacetime.  
We examine multiple radiation models to investigate how
different choices in the emission models affect the 
time-dependent luminosity.
One model assumes only classical 
thermal emission, and a second model incorporates both bremsstrahlung 
and thermal emission according to optical depth.
Finally, we also compare our results
with the simple bremsstrahlung estimate that we obtained in Paper~I.

%
%
\section{Radiation transfer}
\label{sec:rad_transfer}

Electromagnetic radiation follows geodesics of the spacetime. 
As the photon is massless, the radiation follows null geodesics, meaning
that the proper separation between all points on the geodesic is zero.
We calculate the geodesics from the spacetime metric
$g_{\alpha\beta}$,  and use the convention that Greek 
indices take the values $\{0,1,2,3\}$.
The null geodesics are found by solving the first order system
\begin{eqnarray}
\frac{\rmd x^{\mu}}{\rmd\lambda} & = & p^{\mu} \, ,\\
\frac{\rmd p_{\mu}}{\rmd\lambda} & = & g^{\gamma\beta}
                           \Gamma^\alpha{}_{\mu\gamma} p_\alpha p_\beta \, , 
\end{eqnarray}
where $\Gamma^\alpha{}_{\mu\gamma}$ are the
Christoffel symbols, and $x^{\mu}$ and $p^\mu$ are the coordinates and 
momentum of the photon, respectively.
Since we are dealing with null geodesics we parameterize the solution
in terms of the affine parameter $\lambda$.
We solve the geodesic equations using the standard Runge-Kutta fourth
order integration method,
numerically integrating the geodesics backwards in time, perpendicularly from 
a viewing window. This window is set at different locations,
in order to provide different viewing angles of the disk.

After obtaining the geodesic paths in the spacetime, we can now solve
for the electromagnetic radiation intensity along these paths.
The radiation transfer equation for the photon intensity at frequency $\nu$ 
is~\cite{fuerst_wu}
\begin{eqnarray}
\frac{\rmd I_\nu}{\rmd\lambda} 
         = - p^{\alpha} u_{\alpha} \left[ \eta_0 -\chi_0 I_\nu \right],
\label{eqn:rte}
\end{eqnarray}
where $\chi_0$ and $\eta_0$ are, respectively, the absorption and emission
coefficients in the rest frame with respect to the fluid,  and $u_{\alpha}$
is the four-velocity of the fluid.
In vacuum regions, the fluid four-velocity 
corresponds to the four-velocity of coordinate observers.
This first order radiation transfer equation has the
solution~\cite{fuerst_wu}
\begin{eqnarray}
I_\nu(\lambda) & = & I_\nu(\lambda_0) e^{\int_{\lambda_0}^{\lambda} \chi_0(\lambda^\prime,\nu_0) u_\alpha p^\alpha d\lambda^\prime} - \nonumber \\
  & & \int_{\lambda_0}^\lambda e^{\int_{\lambda^\prime}^\lambda \chi_0(\lambda^{\prime\prime},\nu_0)u_\alpha p^\alpha d\lambda^{\prime\prime}} \eta_0(\lambda^\prime,\nu_0) u_\alpha p^\alpha d\lambda^\prime\, , \label{eqn:fuerstwu} 
\end{eqnarray}
where $\nu_0$ is the photon frequency at emission,
and $\lambda_0$ indicates the affine
parameter value where the geodesic begins.   
The integration constant $I_\nu(\lambda_0)$ 
takes into account the photon intensity where the geodesic begins.  
In this work, the
geodesics begin far enough behind the disk so that there
is no matter to act as a source for emission,  and we set
$I_\nu(\lambda_0)$ to be zero.  The 
fluid variables come from a numerical solution that is specified on 
a finite difference grid.  When integrating
Eq.~(\ref{eqn:fuerstwu}), these fluid variables are interpolated
both in space and time.

As mentioned above, we employ two different
emission models in calculating the spectrum from the kicked disk. 
The first model assumes that emission from the optically thin region 
is thermal bremsstrahlung, and the optically thick region is a classical 
blackbody~\cite{Schnittman2006}. The second model uses only thermal
emission.  Both models use Kramer's opacity law for the absorption.
We expect the first model to be more realistic, as the optically thin
regions of the disk are expected to radiate primarily with bremsstrahlung,
rather than thermal emission, and the second model provides a useful
comparison to gauge the importance of the bremsstrahlung emission.
We refer to these models as bremsstrahlung-blackbody and
thermal emission models, respectively.

The bremsstrahlung-blackbody radiation transfer model 
uses the standard bremsstrahlung emissivity and a modified form of 
Kramer's opacity law for the absorption.
The bremsstrahlung emissivity is modeled on a Maxwellian plasma~\cite{tucker}
%
\begin{eqnarray}
\eta_0 & = & 6.8 \times 10^{-38} Z^2 n_e n_i T^{-1/2} \nonumber \\
      & & \times \bar{g}_{\rm ff}(\nu,T) e^{-x} \, 
          \frac{\mbox{erg}}{\mbox{s\, cm$^3$\, Hz}}\, ,
\label{eqn:emission} 
\end{eqnarray}
where 
\begin{equation}
x\equiv h\nu/k T.
\end{equation}
In these equations $h$ is Planck's constant, $k$ is Boltzmann's constant,
$T$ is the temperature of the fluid,
$n_e$ is the electron number density, $n_i$ is the ion
number density, $Z = n_e/n_i$,
and $\bar{g}_{\rm ff}$ is the temperature averaged
Gaunt factor for free-free transitions in pure hydrogen.
The electron and ion number densities are given by
\begin{equation}
n_e  =  \frac{\rho}{\mu_e m_p} \, , \qquad
n_i  =  \frac{\rho}{\mu_i m_p} \, ,
\end{equation}
where
\begin{equation}
\mu_e  =  \frac{2}{1+X} \, , \qquad
\mu_i  =  \frac{4}{1+3X} \, . \label{eqn:mass_fraction} 
\end{equation}
Here $\rho$ is the rest density of the fluid, $m_p$ is the mass of the proton, 
and $X$ is the relative abundance of hydrogen in the universe, which
we set to be  $X=0.75$. 
We use a temperature averaged Gaunt factor, $\bar{g}_{\rm ff}$, based on
\cite{tucker}, where different approximations are given depending on the
values of $T$ and $\nu$. To completely specify $\bar{g}_{\rm ff}$, 
some criteria 
must be used to set the
transition between approximations valid for $\nu\ll kT/h$ and 
$\nu \approx kT/h$ in the regime $T({\rm K}) \gg 3\times10^5 Z^2$. We
set the transition at the values [$\nu$, $T$] for 
which $\bar{g}_{\rm ff}$ in the two approximations coincide, giving
\begin{widetext}
\begin{equation} \label{eq:gaunt}
\bar{g}_{\rm ff}(\nu,T) =  
  \left\{ 
    \begin{array}{c c l | c}
      \displaystyle\frac{\sqrt{3}}{\pi} \left[17.7+\log \left(\frac{T^{3/2}}{\nu Z}\right) \right] 
          & & \quad {\rm if} \quad \nu \leq 10^9 T^{3/2}           &  \\
          & &                                       & \quad {\rm and} \quad T\leq 3 \times 10^5 Z^2 \\
      1
          & & \quad {\rm if} \quad \nu > 10^9 T^{3/2}              &  \\
          & &                                       & \\
\hline
          & &                                       & \\
               \displaystyle x^{2/5} 
          & & \quad {\rm if} \quad  x < 1          &    \\
          & &                                       & \\
      1 
          & & \quad {\rm if} \quad  1 \leq x < 29.6    & \quad {\rm and} \quad  T> 3\times 10^5 Z^2  \\
          & &                                       & \\
      \displaystyle\frac{\sqrt{3}}{\pi} \log \left( 2.2 x\right)
          & & \quad {\rm if} \quad  29.6 \leq x       &   
    \end{array} 
  \right. \quad  
\end{equation}
\end{widetext}
where  $T$ is in units of Kelvin, and $\nu$ in Hertz. 
Following 
\cite{Schnittman2006}, 
we use for the absorption a modified
Kramer's opacity law (compare to Eq.~\ref{eq:kramer} below),
\begin{eqnarray}
\chi_0 = 5 \times 10^{24} \rho^2 T^{-7/2}\left( \frac{1-e^{-x}}{x^3} \right) \,\,\, {\rm cm}^{-1}.
\end{eqnarray} 
This modification has the net effect of adding thermal radiation
for optically thick regions, while optically thin regions radiate
with bremsstrahlung. 
In optically thick regions, all photons are absorbed and 
$\rmd I_\nu/\rmd\lambda \to 0$, 
giving the expected blackbody intensity 
\begin{equation}
I_\nu = \frac{\eta_0}{\chi_0} \propto  T^3 x^3 \frac{e^{-x}}{1-e^{-x}}. 
\label{eqn:blackbody}
\end{equation}

In the second radiation transfer model we consider only the
emission of thermal radiation.  We use  Kramer's opacity law for the absorption
\begin{equation} \label{eq:kramer}
\chi_0 = 5 \times 10^{24} \rho^2 T^{-7/2} \,\,\, {\rm cm}^{-1},
\end{equation}
and specify the emissivity using Kirchoff's law~\cite{Rybicki:1979}
\begin{equation}
\eta_0 = \chi_0 B_\nu(T),
\end{equation}
where $B_\nu(T)$ is given by the Planck law
\begin{equation}
B_\nu(T) =  \frac{2h\nu^3/c^2}{e^x - 1}.
\end{equation}
Thermal radiation is emitted by matter in thermodynamic equilibrium,
while blackbody radiation occurs when the radiation is itself in 
thermodynamic equilibrium.
In optically thick regions ($\rmd I_\nu/\rmd\lambda\to 0$), we again reduce
to the blackbody intensity, $I^{\rm BB}_\nu = B_\nu(T)$.

Having specified our emission models, we briefly mention some physical
effects not yet contemplated in our analysis.
As photons move away from the gravitational sources, their frequencies 
change due to gravitational redshift.  As the redshift varies depending
on the location at which the photon is emitted, this leads to a broadening
of the emission lines.  
A simple estimate in a Schwarzschild spacetime
shows that gravitational redshift accounts for
typical frequency changes of 5\% percent or less if we consider photons being
emitted from a minimum distance to the black holes of about 10 $R_{\rm S}$,
consistent with the disk in our model.
Broadening will occur also due to relativistic Doppler,
especially in the inclined views. Considering the fluid velocities in our
model, up to about $0.28c$, we estimate a maximum Doppler shift of about 25\%.
As the frequency bins used
here are much larger than those changes, we do not include the line
broadening caused by Doppler or gravitational redshift in our model, and we 
work only with the original emission frequencies.

Finally, in Paper~I we estimated the total bremsstrahlung luminosity 
by assuming a transparent gas.  In this approximation the luminosity is 
obtained by integrating the bremsstrahlung emissivity, $\epsilon_{B} \propto \rho^2 T^{1/2}$, over the
computational domain. We compare the results computed here with this earlier
model, and find that this simple estimation works very well
for the higher frequencies that we consider.

%
%

\section{Results}
\label{sec:results}
In paper~I the dynamics of a disk affected by a possible recoil
velocity of a newly merged black hole was studied. A salient feature
of the analysis presented there is that the recoil influences strongly
the disk's behavior as long as it has a component on the orbital plane.
This induces shocks in the disk which in turn heat the gas and can lead
to emissions with a variability tied to the orbital period at the disk.
The results presented in Paper~I were obtained assuming the simple ideal 
fluid equation of state 
\begin{equation}
P = (\Gamma -1)\rho \epsilon,
\label{eq:eos}
\end{equation}
where $P$ is the fluid pressure, $\epsilon$ is the specific internal
energy, and we choose $\Gamma=5/3$, considering the gas as being monatomic.  
We assume that there is no energy transfer from the fluid to the 
electromagnetic radiation, and this equation of state allows for
fluid flows, e.g. shocks, that are not 
isentropic.
To calculate the emissivity and absorption coefficients, we 
obtain the fluid temperature from the ideal fluid law as
\begin{equation}
T = \frac{\mu m_{\rm p}}{k} \frac{P}{\rho},
\end{equation}
where
\begin{equation}
\frac{1}{\mu} = \frac{1}{\mu_e} + \frac{1}{\mu_i},
\end{equation}
and $\mu_e$ and $\mu_i$ are given in 
Eqs.~(\ref{eqn:mass_fraction}).

The simulations in Paper~I correspond to a fixed space-time
background, and thus some variables can be independently rescaled 
a posteriori. For example, one can rescale the black hole's mass and
the disk's density independently, which in turn fixes other 
quantities such as the luminosity. 
This scale invariance, however, does not hold for the emission and
absorption coefficients in the radiation transfer 
equation~(\ref{eqn:fuerstwu}), and we set the physical scales
in the following manner.
We start by setting a central black hole mass $M_{\rm BH} =1 \times 10^8$
$M_\odot$, consistent with super-massive black holes at galactic centers. This
choice fixes the location of the maximum fluid density $\rho_{\rm m}$ in 
our model to $r_{\rm m} \approx 4.5 \times 10^9$~km, approximately 
15 Schwarzschild radii ($R_S$) from the center of the disk. 
To estimate a physically reasonable value for the 
disk's density,  we use an alpha-disk
model~\cite{Shakura-Sunyaev}. Note that this model, unlike ours, 
corresponds to an 
accreting disk. However, as an estimate we assume that the density in
the alpha-disk model will be
similar to our disk at the pre-merger stage, before the inner edge of the disk
``freezes'' and accretion stops
\cite{2002ApJ...567L...9A,Milosavljevic:2004cg}. We set the accretion rate in
the alpha disk model to be an  
Eddington-limited accretion rate of 1 $M_\odot$/year. We choose 
$\alpha=0.1$, a value typical for cataclysmic variables, 
low-mass X-ray binaries, and black hole transients. Using these parameters
in the alpha-disk model we obtain a density 
$\rho \approx 0.01$~g/cm$^3$ at $r=r_{\rm m}$. Hence, we scale the disk
density so that $\rho_{\rm m}=0.01$~g/cm$^3$.

As discussed in Paper~I, the qualitative behavior for all of the
disks that we considered was similar, provided that the recoil
velocity has a component orthogonal to the disk's angular momentum. 
We thus adopt as a representative example the data
obtained for a kick perpendicular to the axis of rotation at 3000 km/s.
We place a square-shaped window at three different locations, providing viewing
angles of 
$0^\circ$, $60^\circ$, and $75^\circ$ with respect to the disk's rotational
axis, while the azimuthal direction is chosen perpendicular to the kick.
In all cases the window is located at a distance of $1.05 \times 10^{10}$km
($\simeq 35 R_S$) away from the black hole (at the 
edge of the computational domain employed in Paper~I). 
The area of the observation window was $4.41 \times 10^{20}$ {km}$^2$ so that the whole
of the disk could be analyzed in such window.
The number of geodesics used for the window and the number of points used to integrate the
radiation transfer equation were selected so that the results did not show appreciable differences
upon further increasing them.

We concentrate on photons having initial frequencies from infrared to
gamma ray (within the range $10^{12}$--$10^{21}$~Hz) and  
integrate the radiation transfer equation
 to the window where we examine the resulting luminosity
vs initial frequency.  We compare results obtained from
the bremsstrahlung-blackbody and thermal models using radiation transfer.
We also compare bremsstrahlung luminosity estimated in the 
transparent limit (Paper~I) with the results from radiation transfer. This allows us to examine interesting
frequency dependencies not highlighted by the bremsstrahlung luminosity estimate by itself.

\begin{figure}[ht]
\begin{center}
\includegraphics[angle=270, width=0.9\columnwidth,clip]{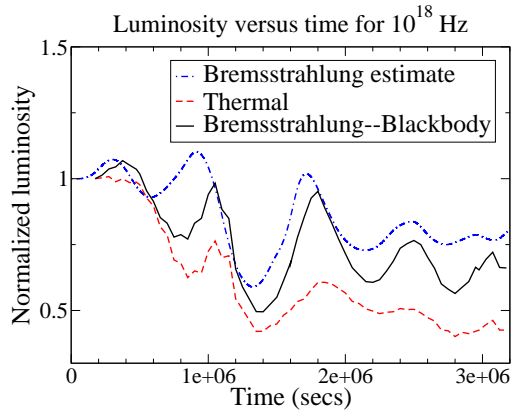}
\caption{Normalized luminosity as a function
of time using different emission models for a frequency of $10^{18}$~Hz.
The horizontal axis shows time in seconds after the kick.
In all cases, the results are normalized with respect to their initial value.
The curve labeled ``bremsstrahlung estimate'' (dash-dot line) is calculated by assuming
a transparent gas
and integrating $\rho^2 T^{1/2}$ over the computational domain.
The curve labeled ``thermal''  (dashed line) is obtained 
using the thermal radiation model described in the text
for a $0^\circ$ viewing angle (top view).
The one labeled ``bremsstrahlung-blackbody'' (solid line) is obtained by using the
bremsstrahlung-blackbody radiation model described in the text,
also at a $0^\circ$ viewing angle.  
  \label{f:1E18_time}}
\end{center}
\end{figure}

\begin{figure}[ht]
\begin{center}
\includegraphics[angle=270, width=0.9\columnwidth,clip]{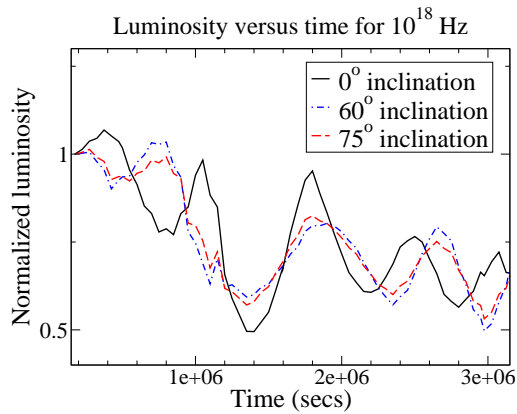}
\caption{Normalized bremsstrahlung-blackbody luminosity as a function
of time for different viewing angles for a frequency of 
$10^{18}$~Hz.
The horizontal axis is time in seconds after the kick.
  \label{f:1E18_timeB}}
\end{center}
\end{figure}

For instance, Figures~\ref{f:1E18_time} and \ref{f:1E18_timeB} illustrate the luminosity
for the case of soft x-ray photons (initial frequency of $10^{18}$~Hz).  Figure~\ref{f:1E18_time}
shows normalized luminosity curves (with respect to the initial value) obtained
via the radiation transfer equation for a window located at $0^\circ$ viewing
angle (the top view) while Figure~\ref{f:1E18_timeB} shows normalized luminosity curves 
for different viewing angles.  In Figure~\ref{f:1E18_time} the results for two radiation models are shown:
bremsstrahlung for emissivity and modified Kramer's opacity law for 
absorption (bremsstrahlung-blackbody),
and thermal emissivity with unmodified Kramer's opacity law (thermal).
Also included is the normalized bremsstrahlung luminosity from Paper~I
which
was calculated assuming a transparent gas, and integrated over all frequencies
(see Eq.~16 of Paper~I).  The good agreement seen in Figure~\ref{f:1E18_time} 
between the bremsstrahlung-blackbody
radiation transfer luminosity curve and the bremsstrahlung luminosity estimate from Paper~I
indicates that the disk is largely transparent at $10^{18}$~Hz.

\begin{figure}[ht]
\begin{center}
\includegraphics[angle=270, width=0.9\columnwidth,clip]{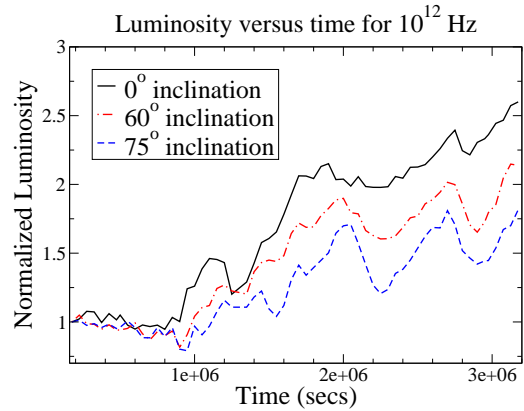}
\caption{Normalized luminosity as a function
of time for different viewing angles for a frequency of $10^{12}$~Hz.
The luminosity was normalized with respect to its initial value.
The horizontal axis labels time in seconds after the kick. 
At $10^{12}$~Hz much of the disk is opaque
and exhibits self-eclipsing behavior. 
The oscillations seen after $t=1 \times 10^6$ seconds are tied to the orbital period of the
disk and are more pronounced as the viewing angle increases.
  \label{f:1E12_time}}
\end{center}
\end{figure}

\begin{figure}[ht]
\begin{center}
\includegraphics[angle=270, width=0.9\columnwidth]{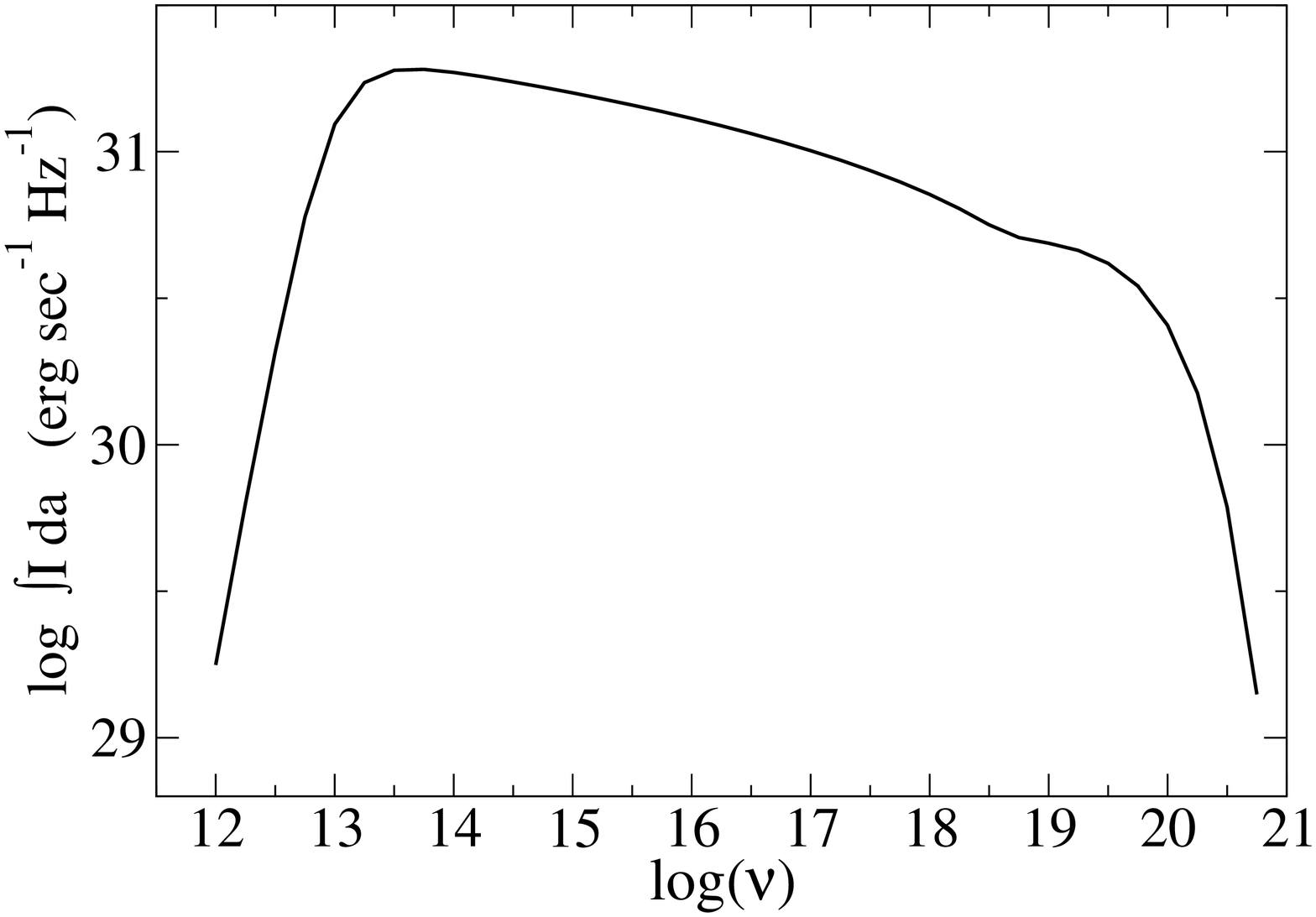}
\caption{
The electromagnetic spectrum computed $1.25 \times 10^6$ seconds
after the kick using the ``bremsstrahlung-blackbody'' radiation model at a
$0^\circ$ viewing angle.
The horizontal axis is the $\log_{10}$ of the frequency.  The vertical 
axis is the $\log_{10}$ of the intensity integrated over the window area.  
In the lower frequencies of the spectrum $~10^{12}-10^{13}$~Hz we see
the expected Rayleigh-Jeans $\nu^2$ frequency dependence.  The small
kink in the spectrum in between $10^{18}$ and $10^{19}$~Hz is due
to the behavior of the Gaunt factor in that region.  
The spectrum was calculated using  36
frequencies, spaced evenly in log space from
$10^{12}$--$5.6\times10^{21}$~Hz.
}
\label{f:spectrum}
\end{center}
\end{figure}

Interestingly this behavior no longer holds for lower frequencies. For instance, 
Figure ~\ref{f:1E12_time} illustrates the results obtained in the infrared 
(initial frequency of  $10^{12}$~Hz).
At this lower frequency, much of the disk is opaque and a self-eclipsing behavior
is increasingly evident as the inclination angle increases. More importantly, after
the transient behavior, a strong increase in relative luminosity is observed as 
shocks significantly affect the dynamics of the disk.  
Finally, Fig.~\ref{f:spectrum} shows the electromagnetic spectrum calculated
from 36 frequencies in the range from $10^{12}$--$5\times 10^{21}$~Hz.

As an illustration of the disk structure observed at different initial frequencies
and viewing angles, Fig.~\ref{f:intensity2} and Fig.~\ref{f:intensity3}  
illustrate
the intensity obtained at different times. The color map shows areas of higher
intensity emission. Comparing these two figures one can clearly see the
differences between a mostly transparent case (Fig.~\ref{f:intensity2}), and
one with absorption (Fig.~\ref{f:intensity3}), showing self eclipsing, which
supports the oscillations shown in Fig.~\ref{f:1E12_time}, seen only
in the opaque case.

%
%
\section{Final Comments}
We have analyzed the possible emission of a disk perturbed by
a recoiling black hole by directly imaging the disk with
radiation transfer.
For frequencies above $\sim 10^{13}$~Hz our studies reveal a characteristic
luminosity variability that could be present in observations of this kind
of systems. This variability shows the same main features as those in our previous results, in which a
simpler model --without radiation transfer-- was used to estimate the
luminosity. 

For frequencies below $\sim 10^{13}$~Hz
the spectrum matches the Rayleigh-Jeans frequency dependence
expected from a blackbody.
Furthermore, at these lower frequencies, a self-eclipsing behavior at
lower frequencies is clearly seen as the disk orbits around the black hole, so that bright and
opaque regions alternatively superpose along the line of sight,
resulting in an induced variability tightly tied to the orbital motion and
inclination. This effect is less pronounced as higher frequencies
are considered and so it is unlikely this feature will be observed. 

Nevertheless, a strong overall variability is observed at all frequencies
associated with shock heating in sectors of the disk which could be detected
provided that the increase in the disk's luminosity prior to the recoil is sufficiently low.
This appears as a likely scenario as the binary hollows out the inner regions of the
disk~\cite{Milosavljevic:2004cg}.
One should be cautious that uncertainties on the details of these disks
and emission processes can affect the characteristics of the luminosities
obtained. In particular, the work of~\cite{Corrales:2009nv}, which employs
a different equation of state, displays interesting differences 
with the ones presented here.
However, regardless of particular
differences induced by details of the equation of state employed,
a common message of all
related works~\cite{kickdisk1,kickdisk2,massloss,kickdisk,Corrales:2009nv} is
that emissions from circumbinary disks around merging black holes possibly
display particular features that could be detected.
These features would follow the black hole merger in the time-frame of months and
would require a good localization in the sky to follow gravitational wave observations
with these electromagnetic counterparts. As already discussed in the literature
(e.g.~\cite{Bloom:2009vx} and references cited therein) LISA's localization will unlikely be
able to single out a preferred galaxy  and so prospects of detection will be greatly
enhanced if pre-cursors or prompt emissions in the electromagnetic wave band are also
available~\cite{Bloom:2009vx}. In the case of binary black hole mergers 
that give
rise to the scenario analyzed in this work, such a possibility is just beginning to be analyzed in
detail~\cite{Palenzuela:2009yr,chang,vanMeter:2009gu,palenzuelainprepBBHEM,moestainprepBBHEM}.

%
%
\noindent{\bf{\em Acknowledgments:}}
We would like to thank
J. Frank, C. Palenzuela, E. Hirschmann, P. Motl, 
S. Liebling, J. Staff and J. Tohline for
 stimulating  discussions.
This work was supported by the NSF grants
PHY-0803629 to LSU,  PHY-0803615 and CCF-0832966
to BYU. Computations were done at BYU,
LONI, LSU, and TeraGrid.

%
%
\bibliography{./bhbhem}
\bibliographystyle{apsrev}

\begin{widetext}

\begin{figure}[h!]
\begin{center}
$\begin{array}{c@{\hspace{.1in}}c@{\hspace{.1in}}c}
\includegraphics[angle=270,width=2.2in]{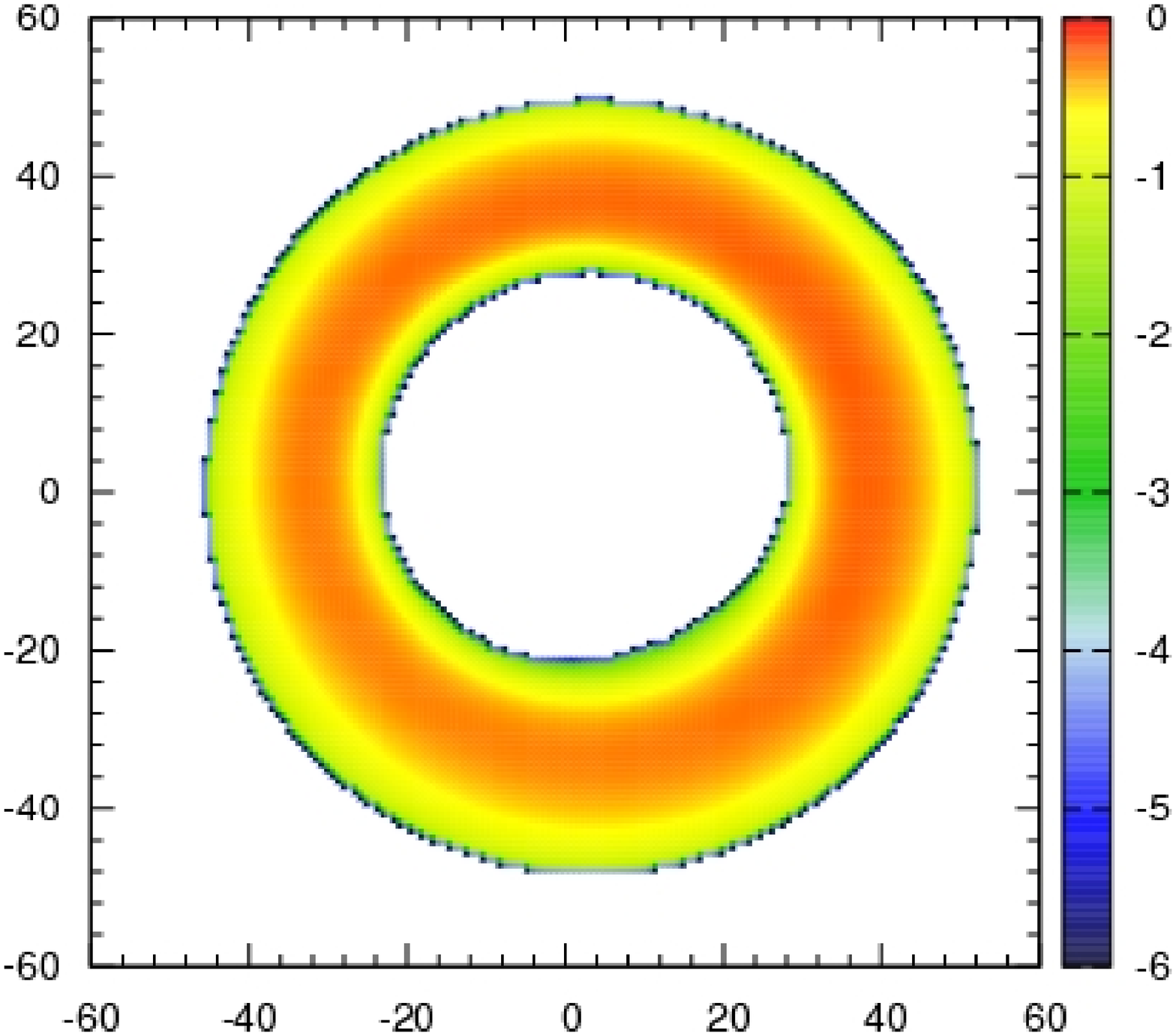}  &  
\includegraphics[angle=270,width=2.2in]{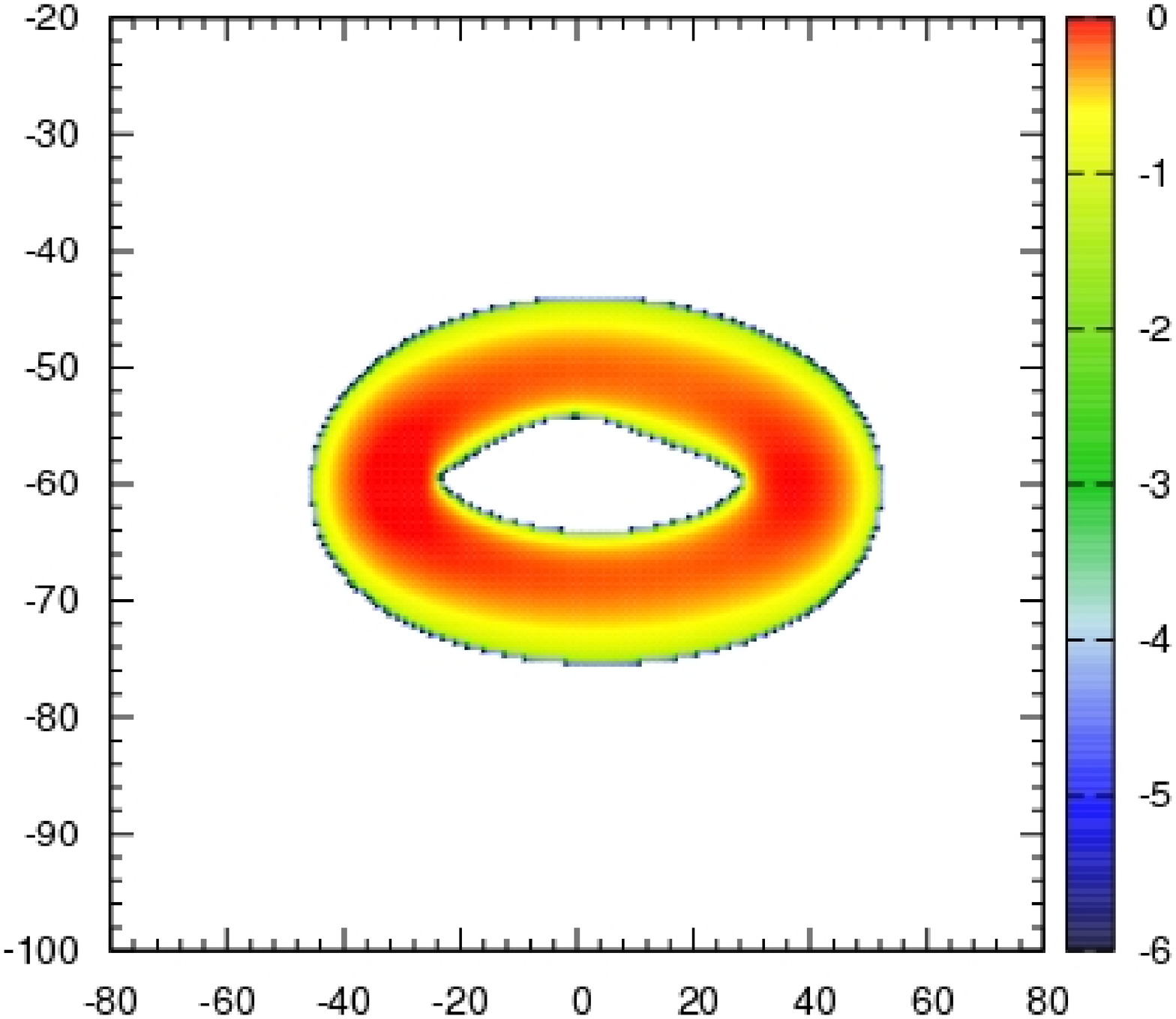}  & 
\includegraphics[angle=270,width=2.2in]{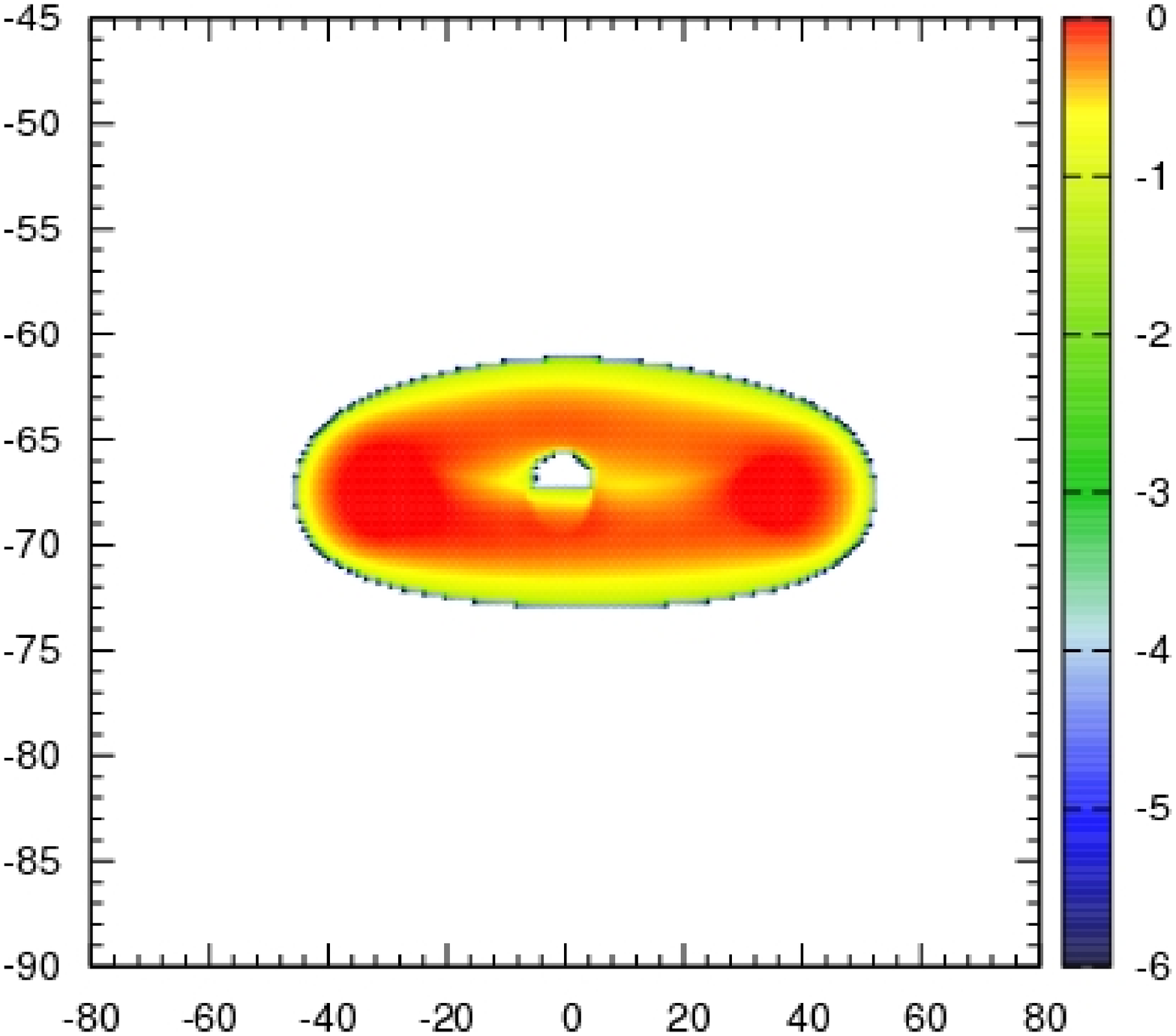} \\
\includegraphics[angle=270,width=2.2in]{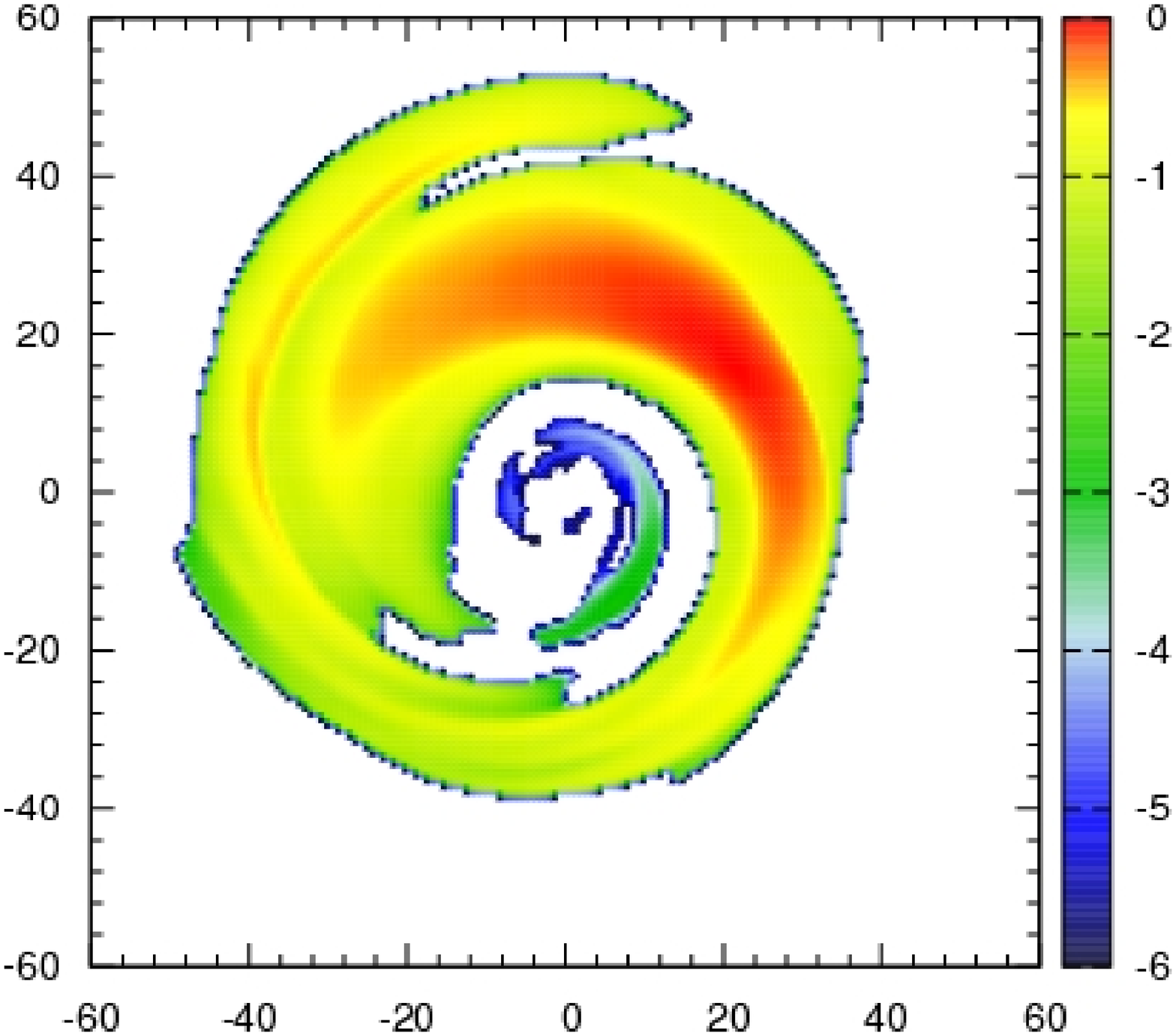}  &  
\includegraphics[angle=270,width=2.2in]{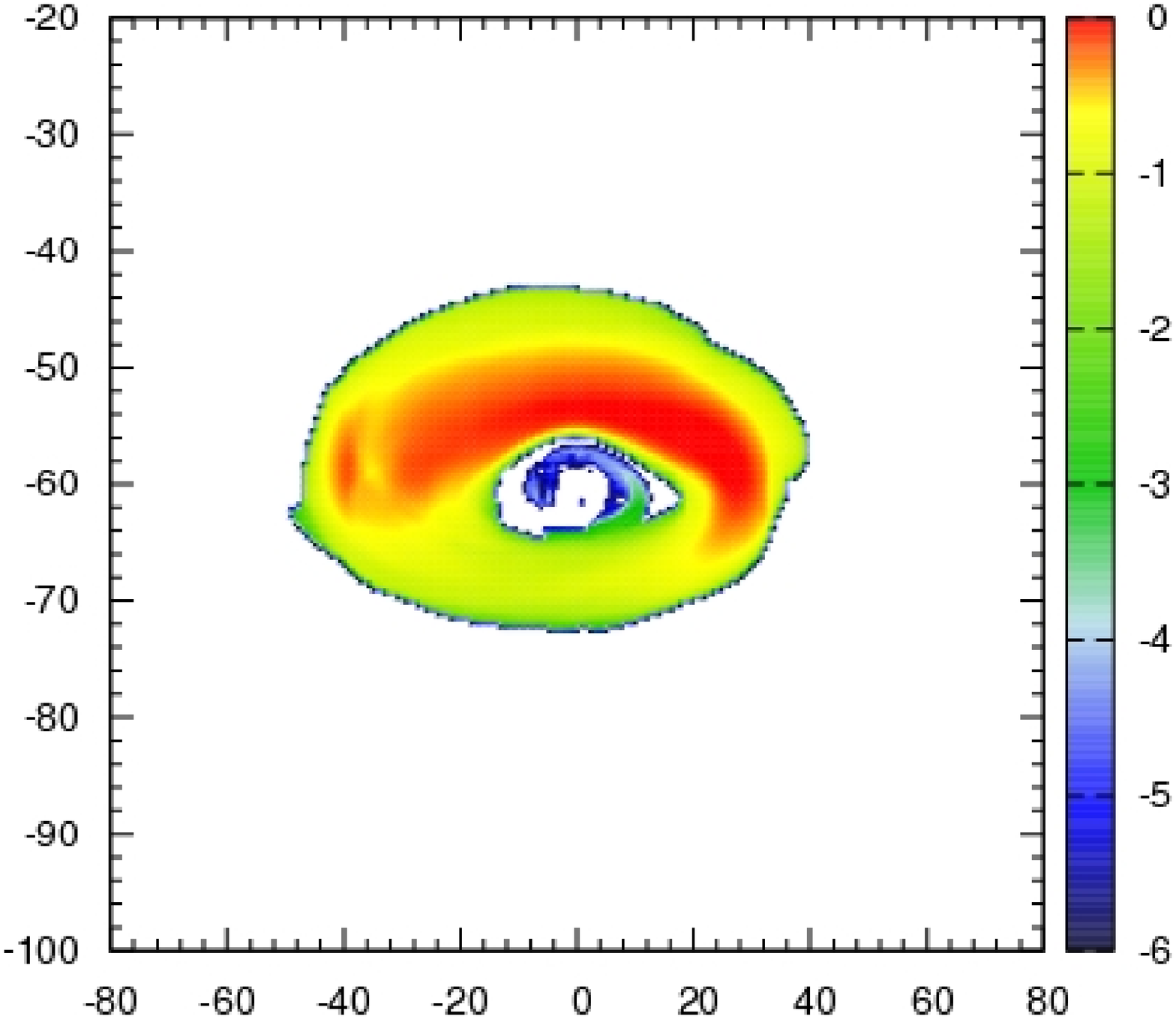}  & 
\includegraphics[angle=270,width=2.2in]{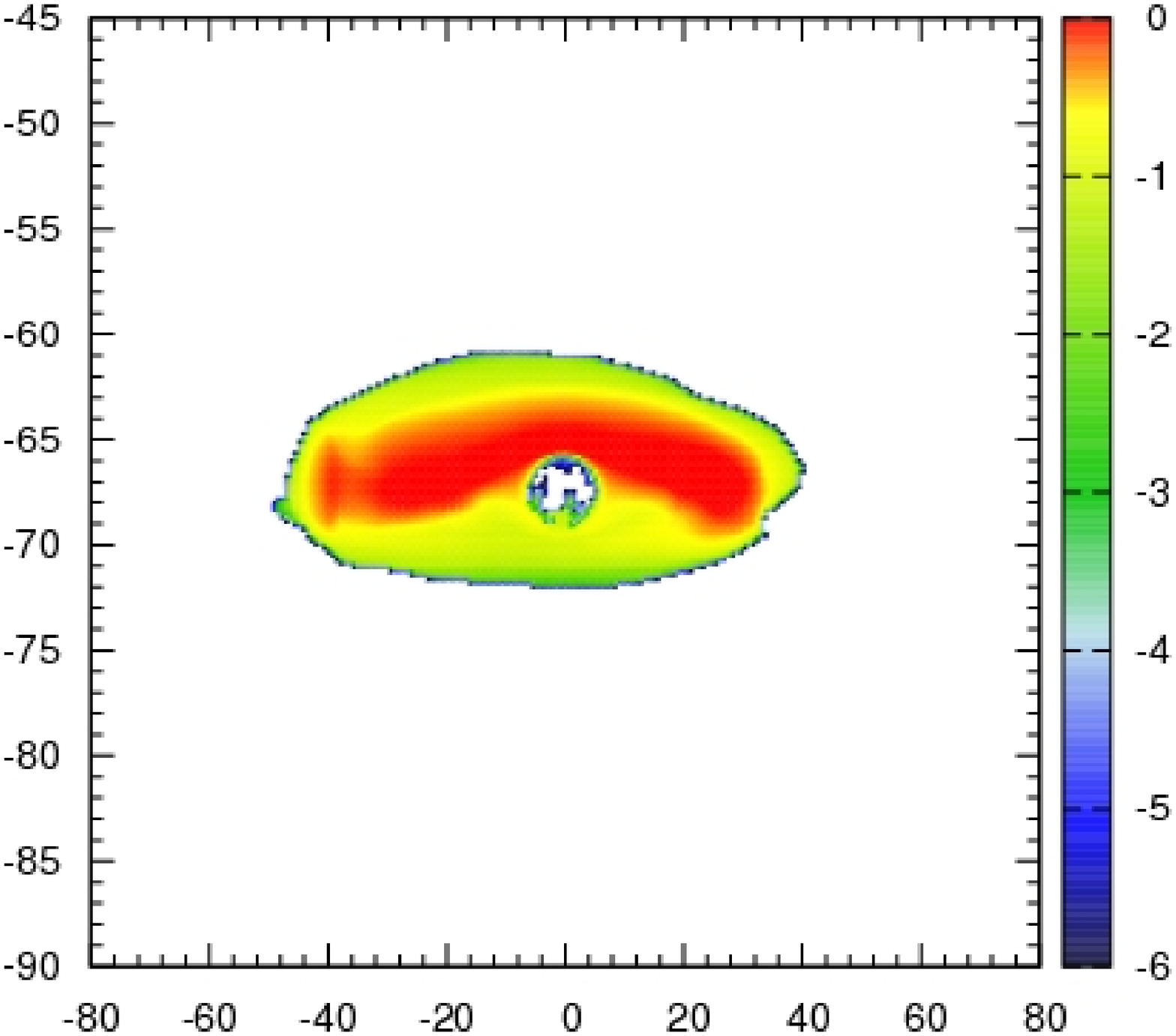} \\
\includegraphics[angle=270,width=2.2in]{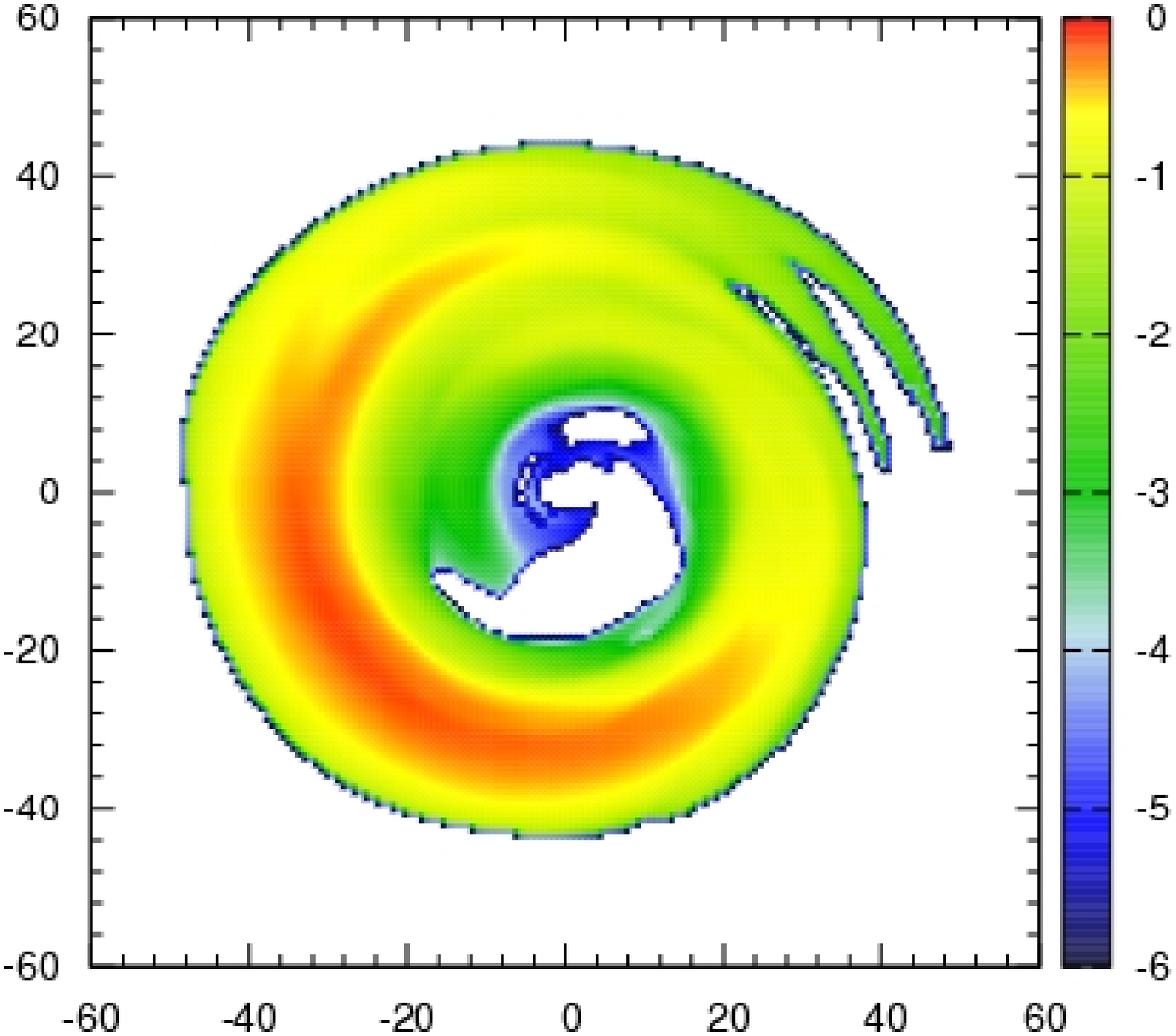} &
\includegraphics[angle=270,width=2.2in]{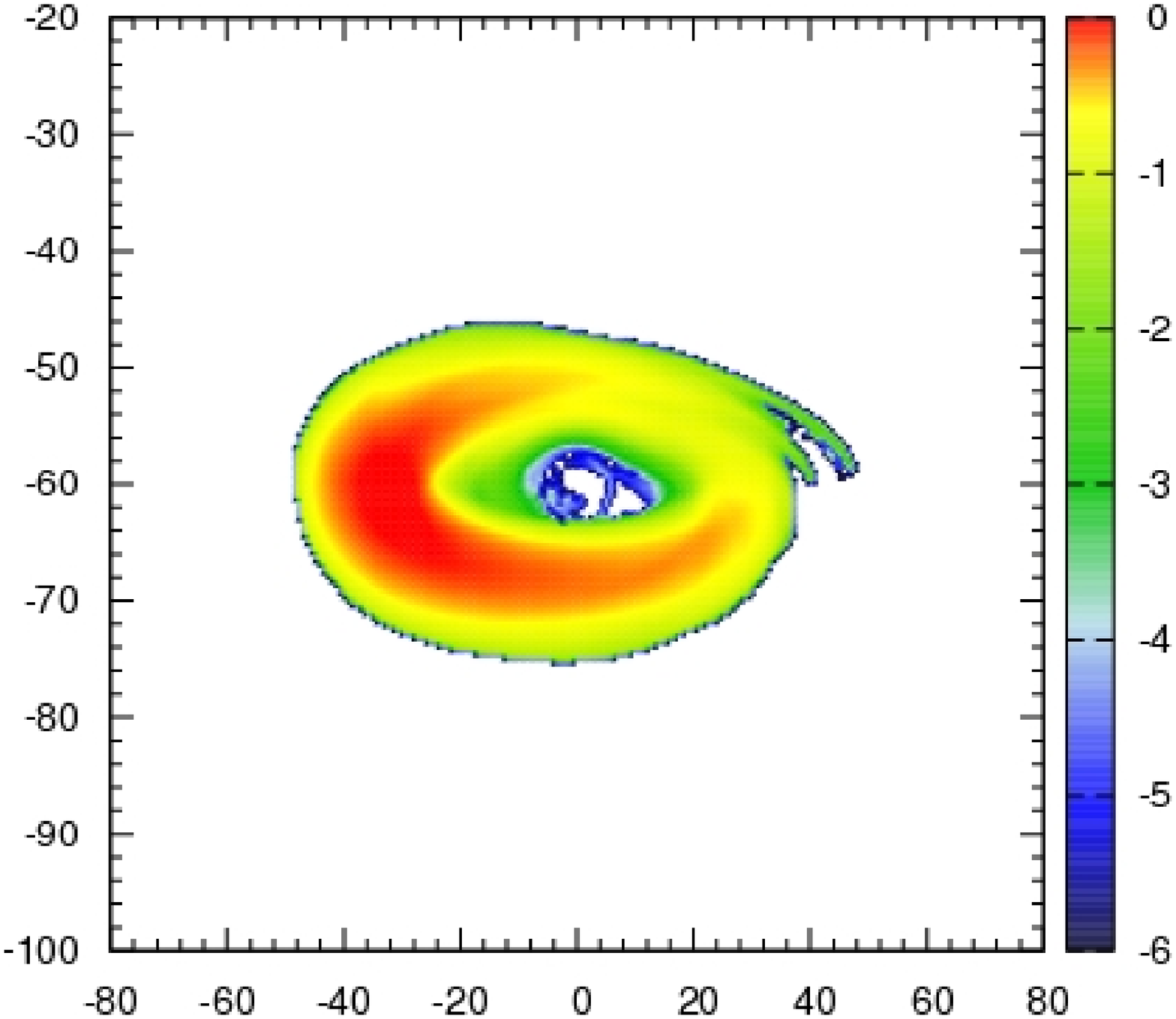} & 
\includegraphics[angle=270,width=2.2in]{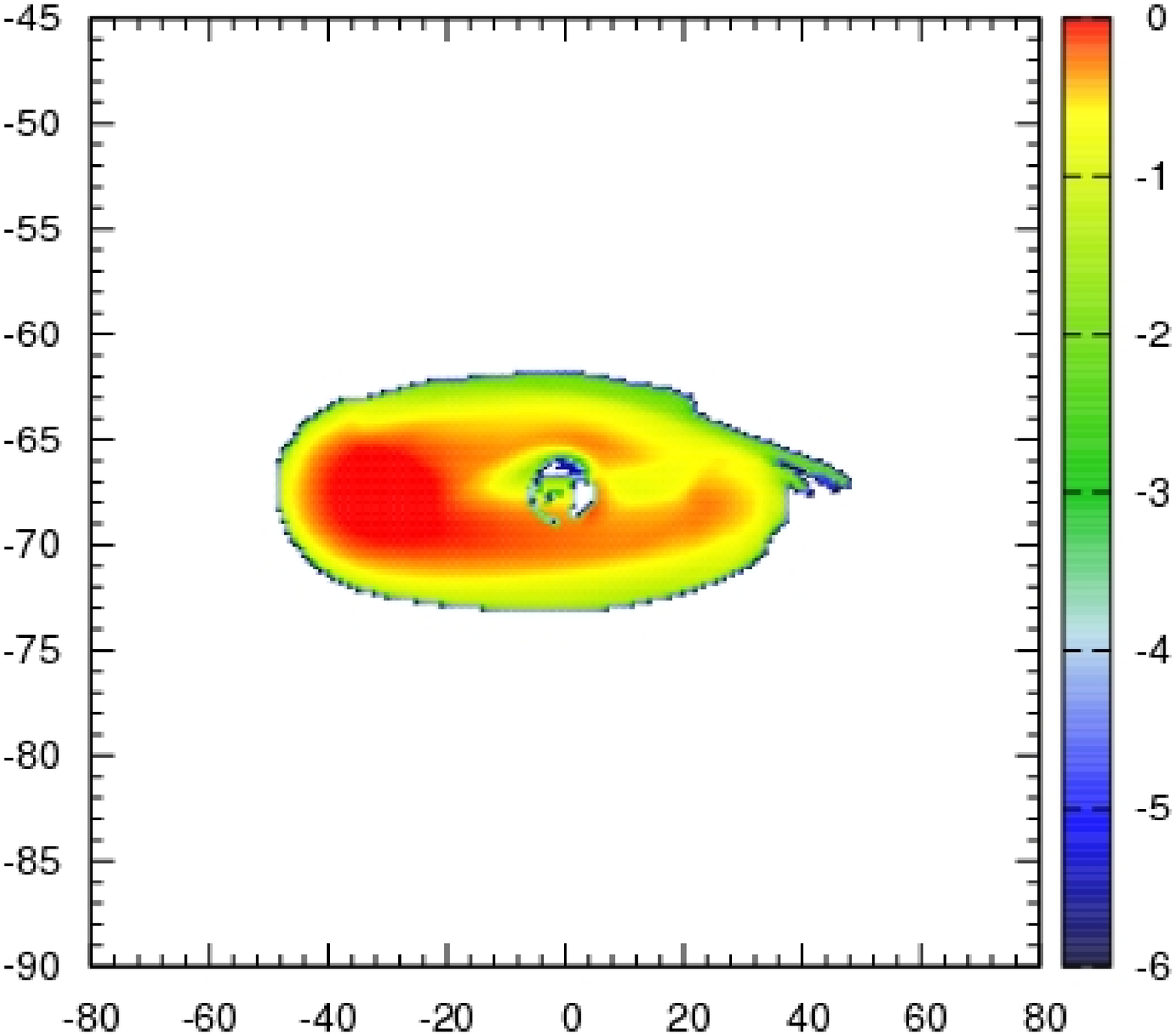} \\
\includegraphics[angle=270,width=2.2in]{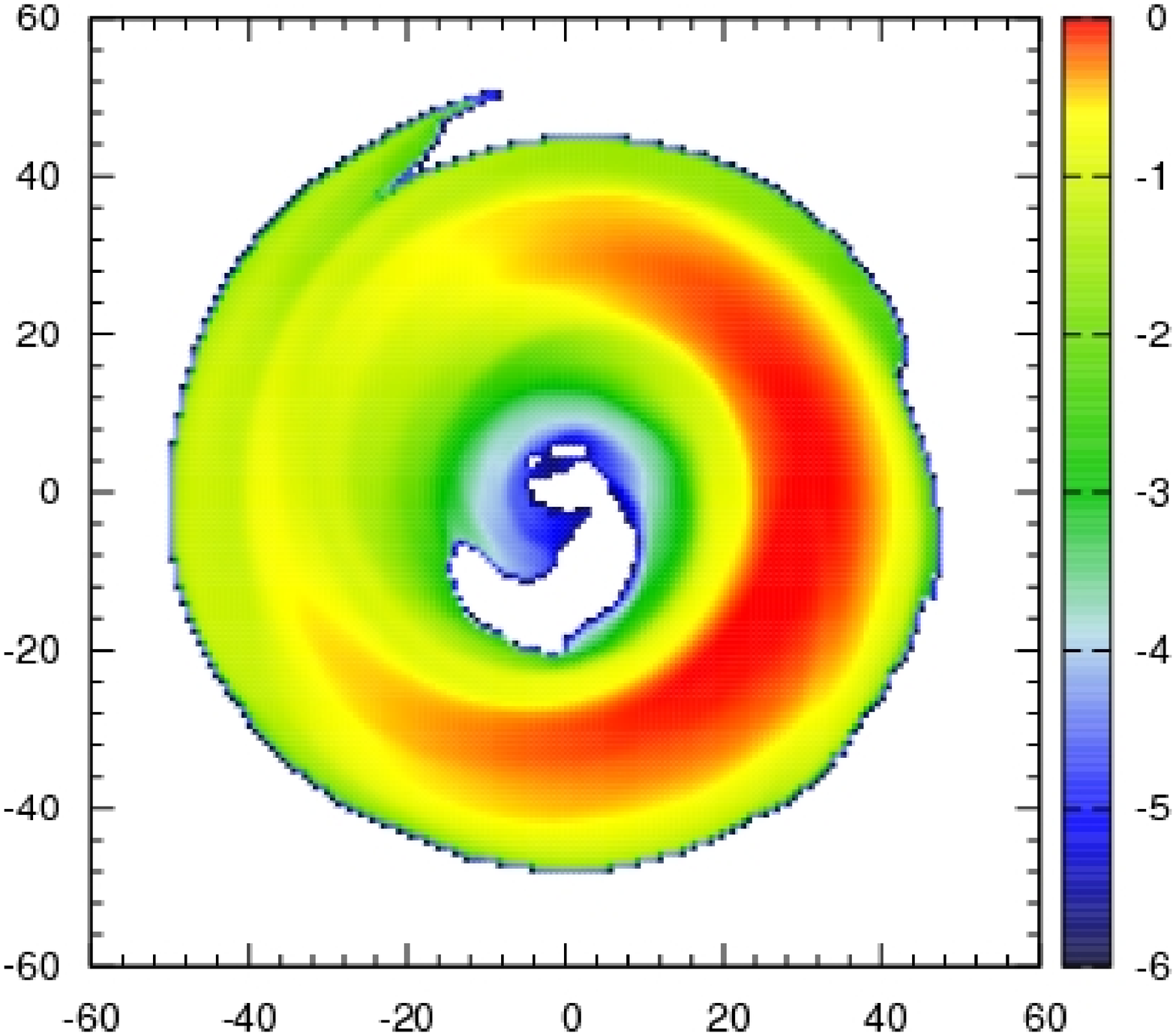} & 
\includegraphics[angle=270,width=2.2in]{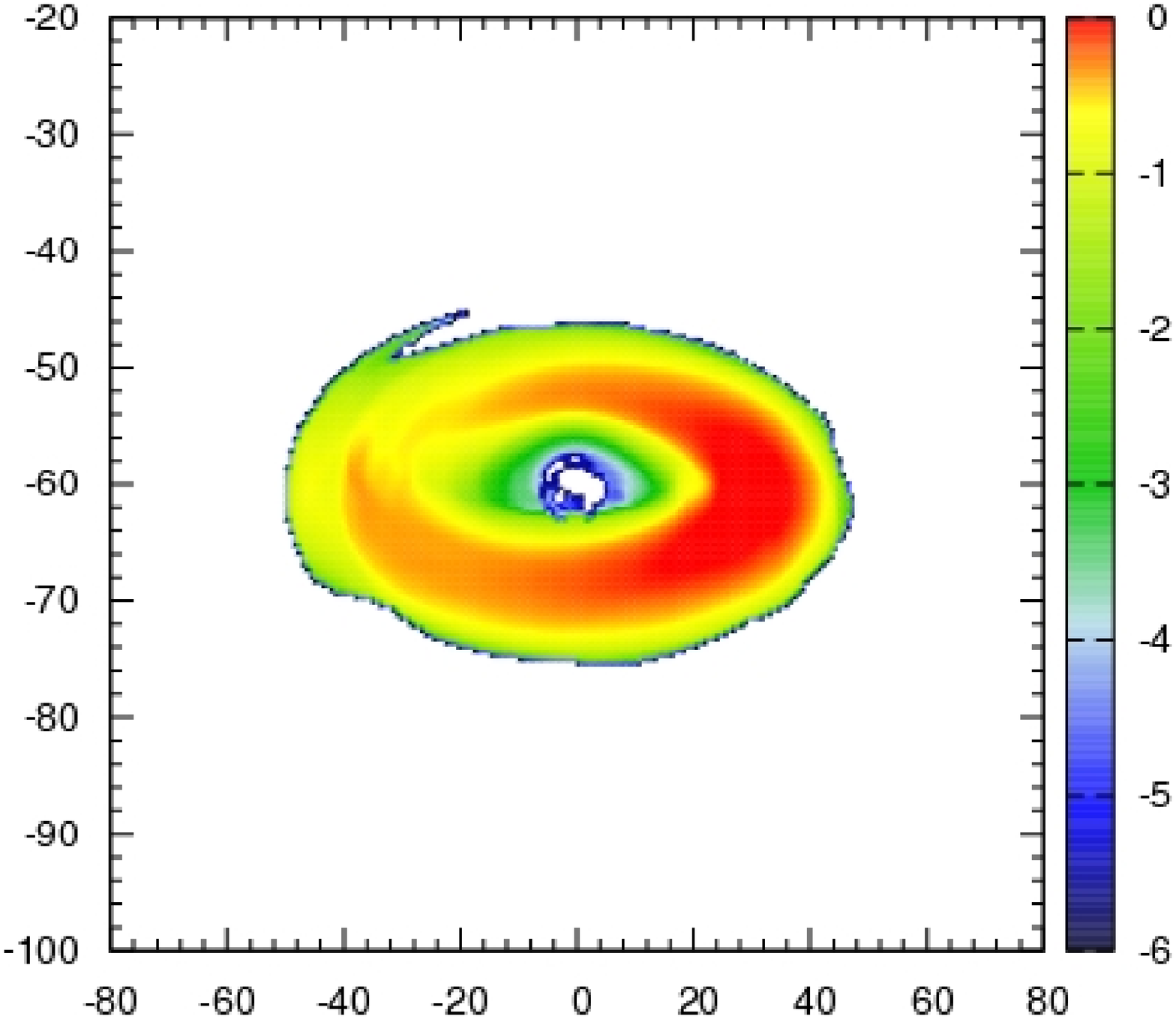} & 
\includegraphics[angle=270,width=2.2in]{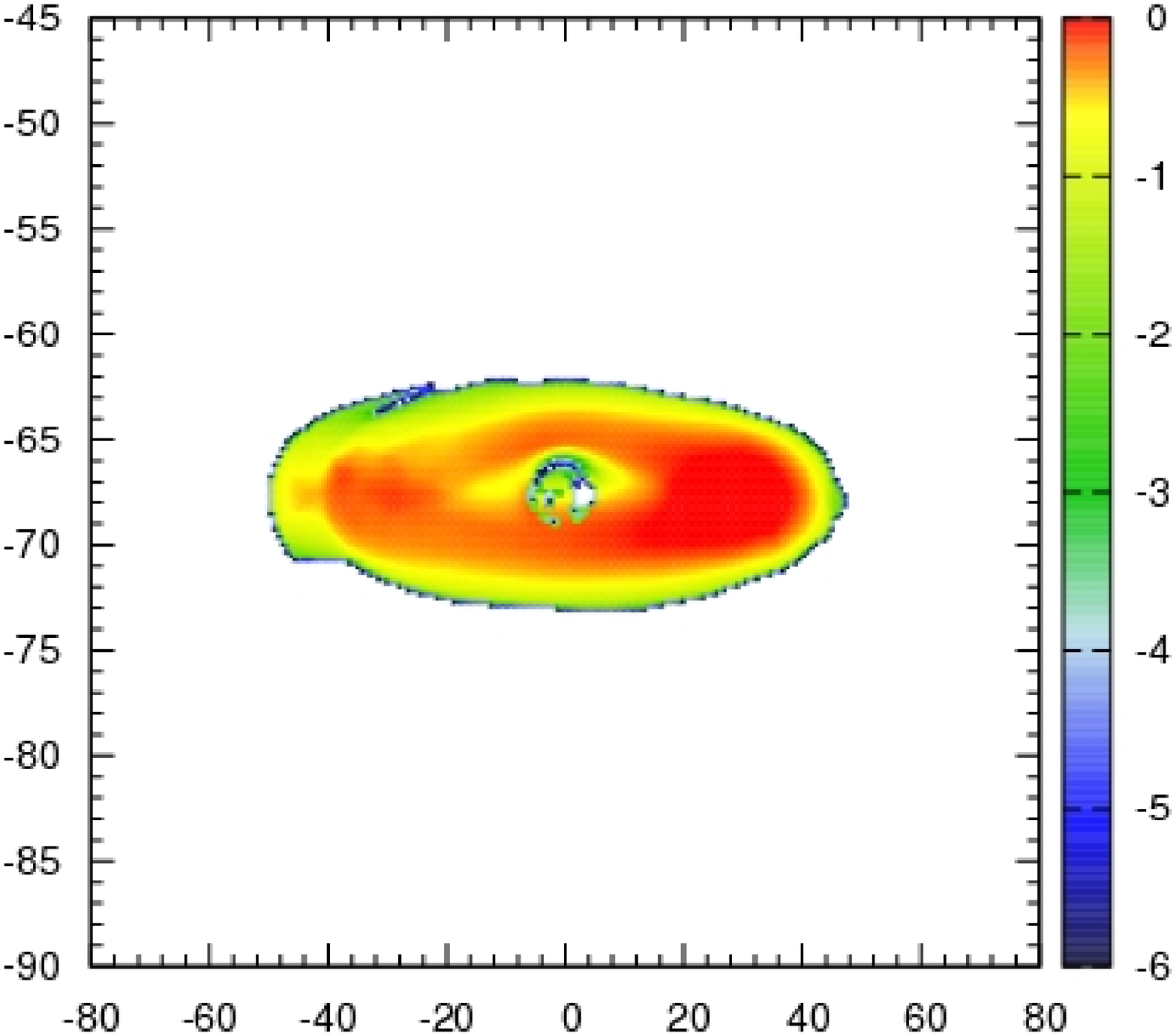}   \\
\end{array}$
\end{center}
\caption{The intensity for $10^{18}$~Hz shown, from top to bottom, 
at $t=1.75 \times 10^5, 1.25\times 10^6, 1.5 \times 10^6,$ and $1.75 \times 10^6$ seconds
after the kick using
the ``bremsstrahlung-blackbody'' radiation model.  The colormap is a log scale;
The intensity for all the images has been normalized using
the maximum intensity found at $t=1.25\times 10^6$ seconds after the kick.
The images on the left/center/right columns correspond to viewing angles of 
$0^\circ$/$60^\circ$/$75^\circ$ respectively.
The number of geodesics used was 40,000.
The axis scale of the images is in units of mass of the black hole.
\label{f:intensity2}}
\end{figure}

\begin{figure}[h!]
\begin{center}
$\begin{array}{c@{\hspace{.1in}}c@{\hspace{.1in}}c}
\includegraphics[angle=270,width=2.2in]{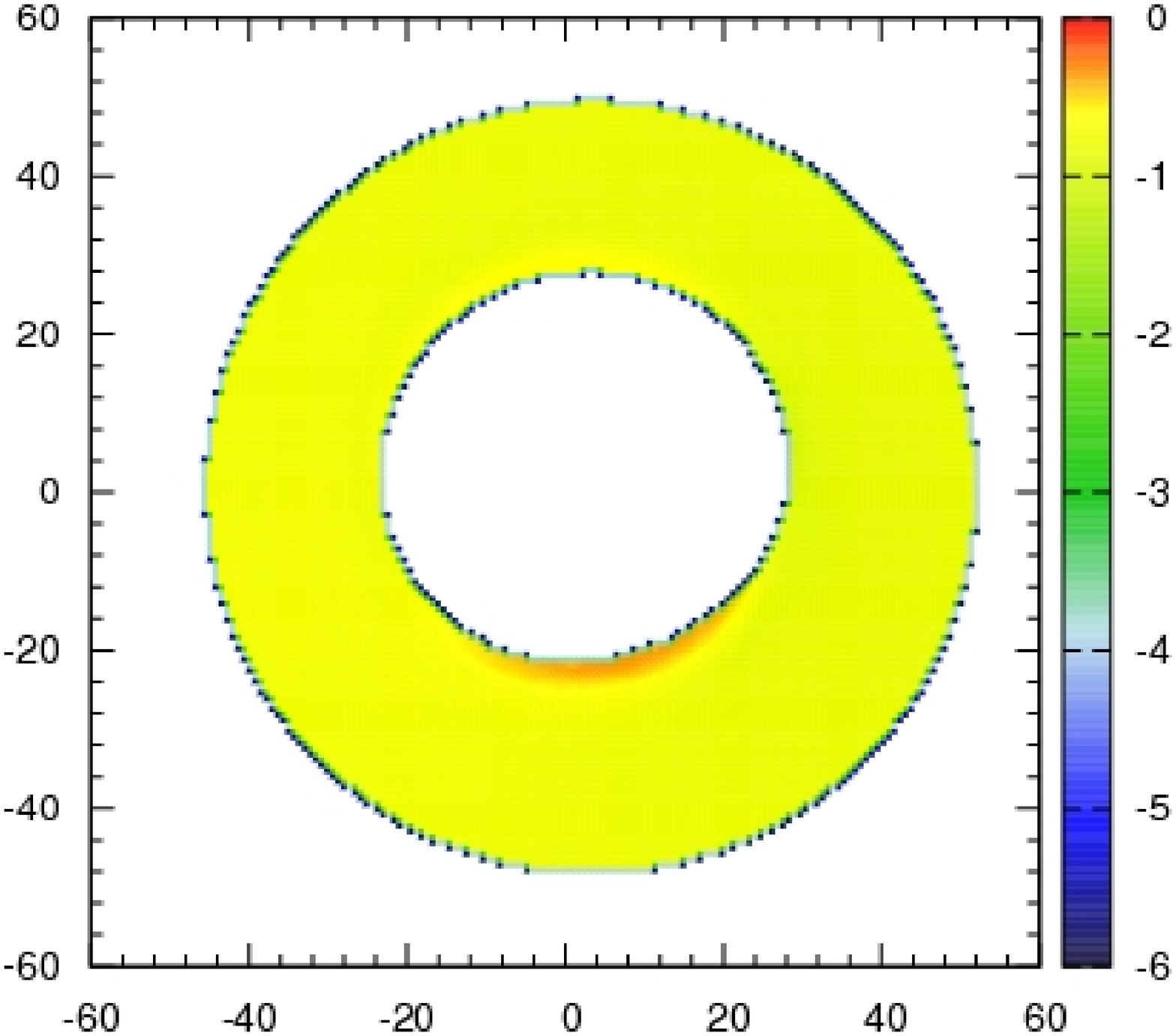}  &  
\includegraphics[angle=270,width=2.2in]{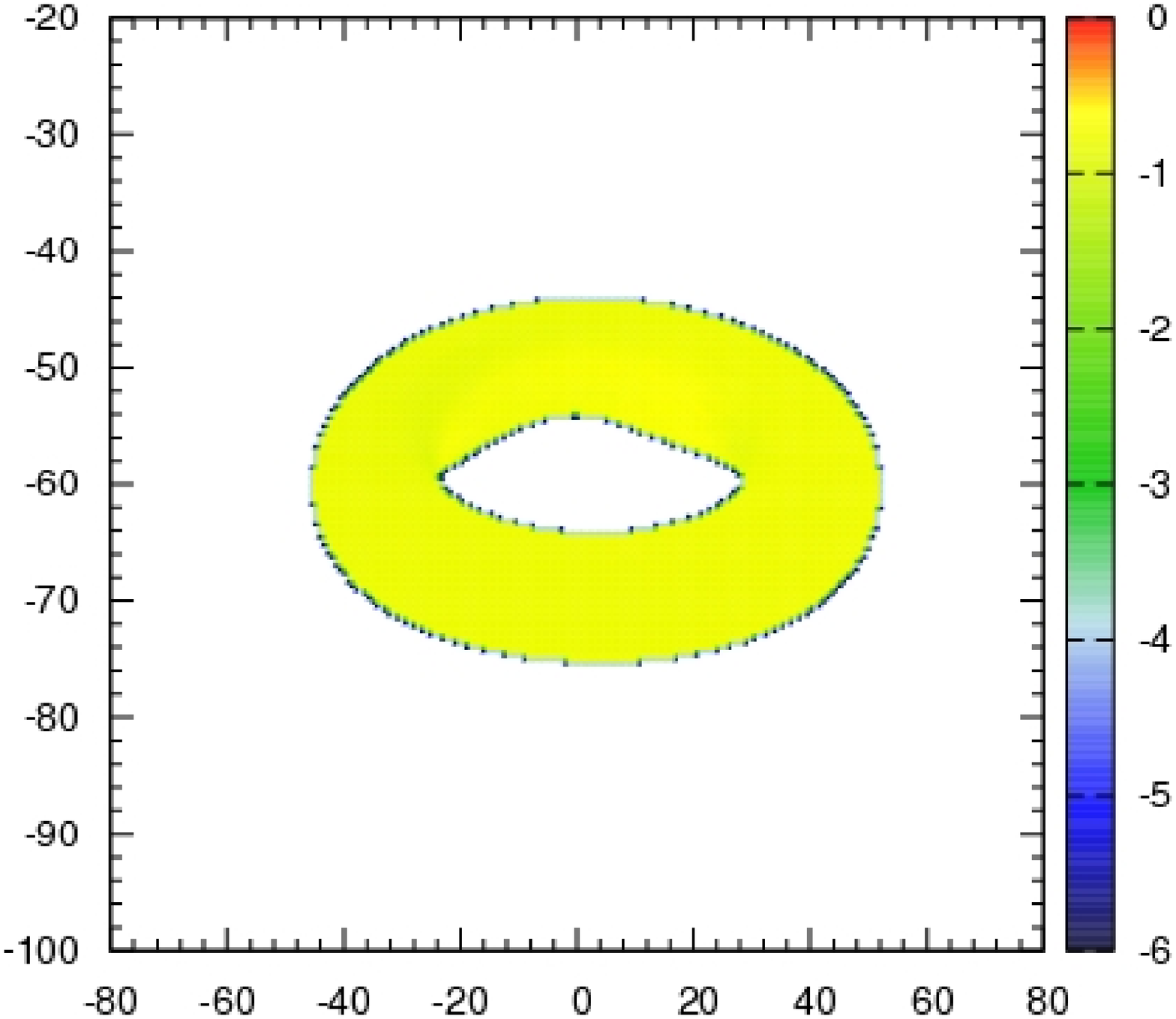}  &
\includegraphics[angle=270,width=2.2in]{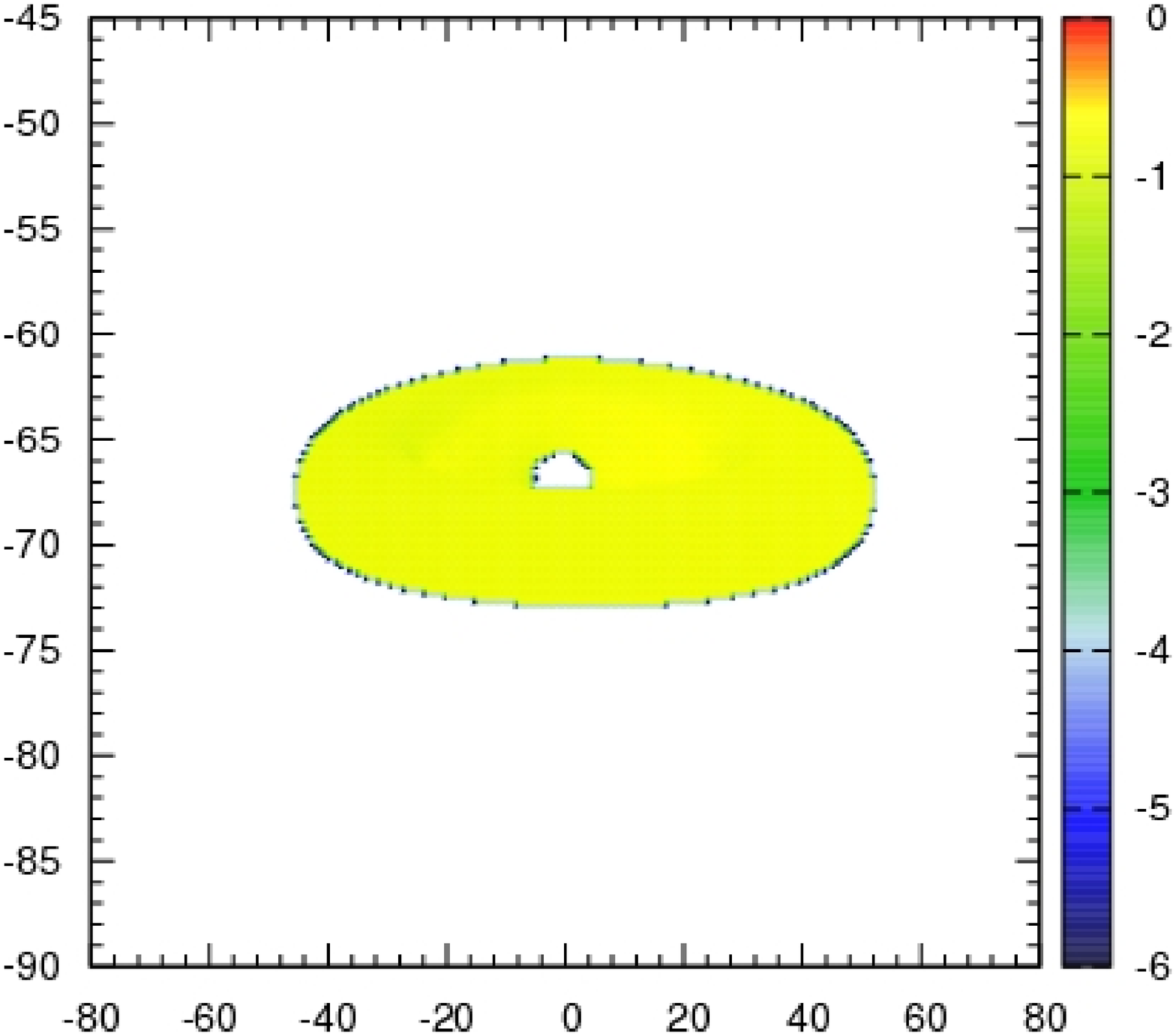}  \\
\includegraphics[angle=270,width=2.2in]{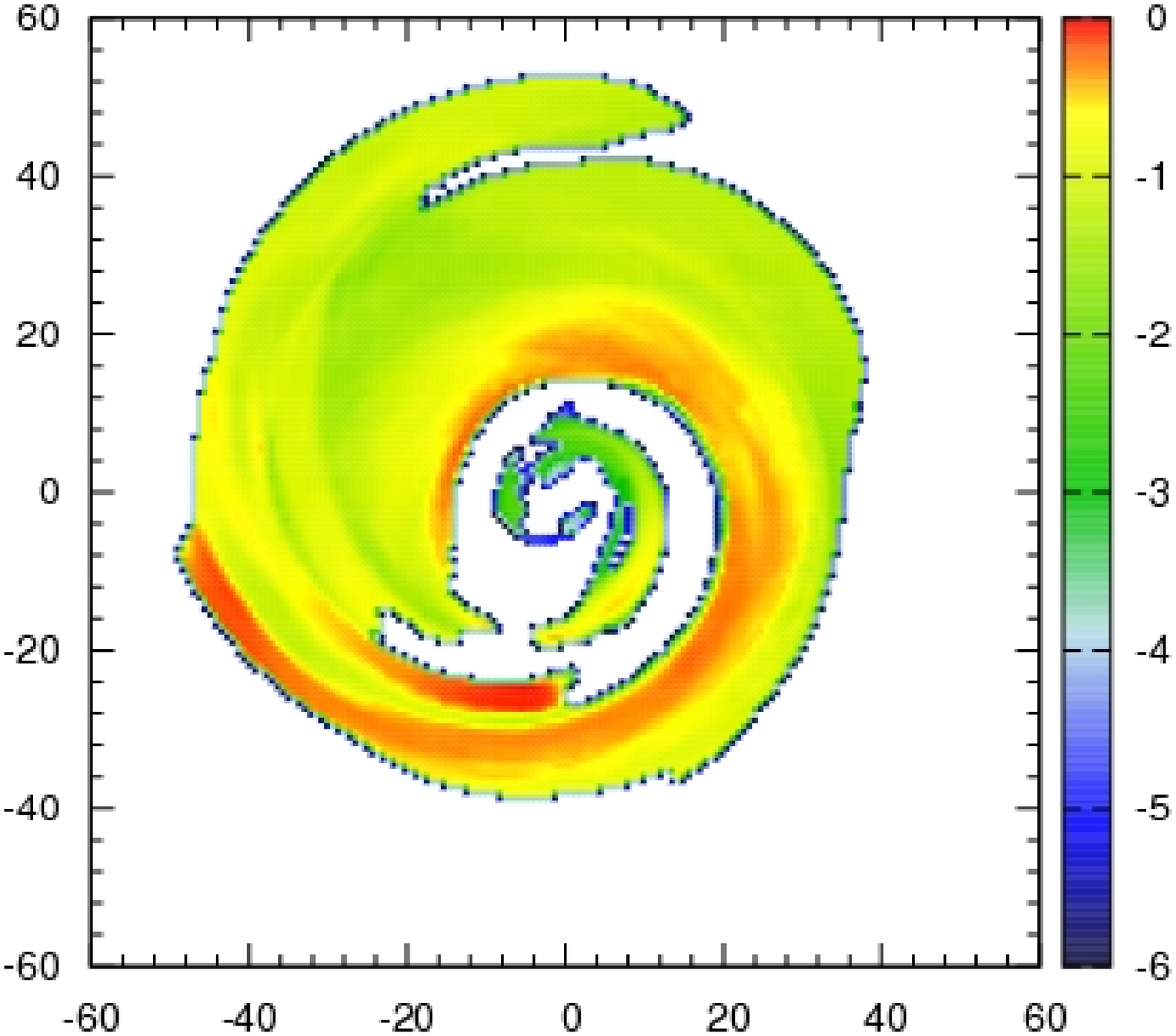}  &  
\includegraphics[angle=270,width=2.2in]{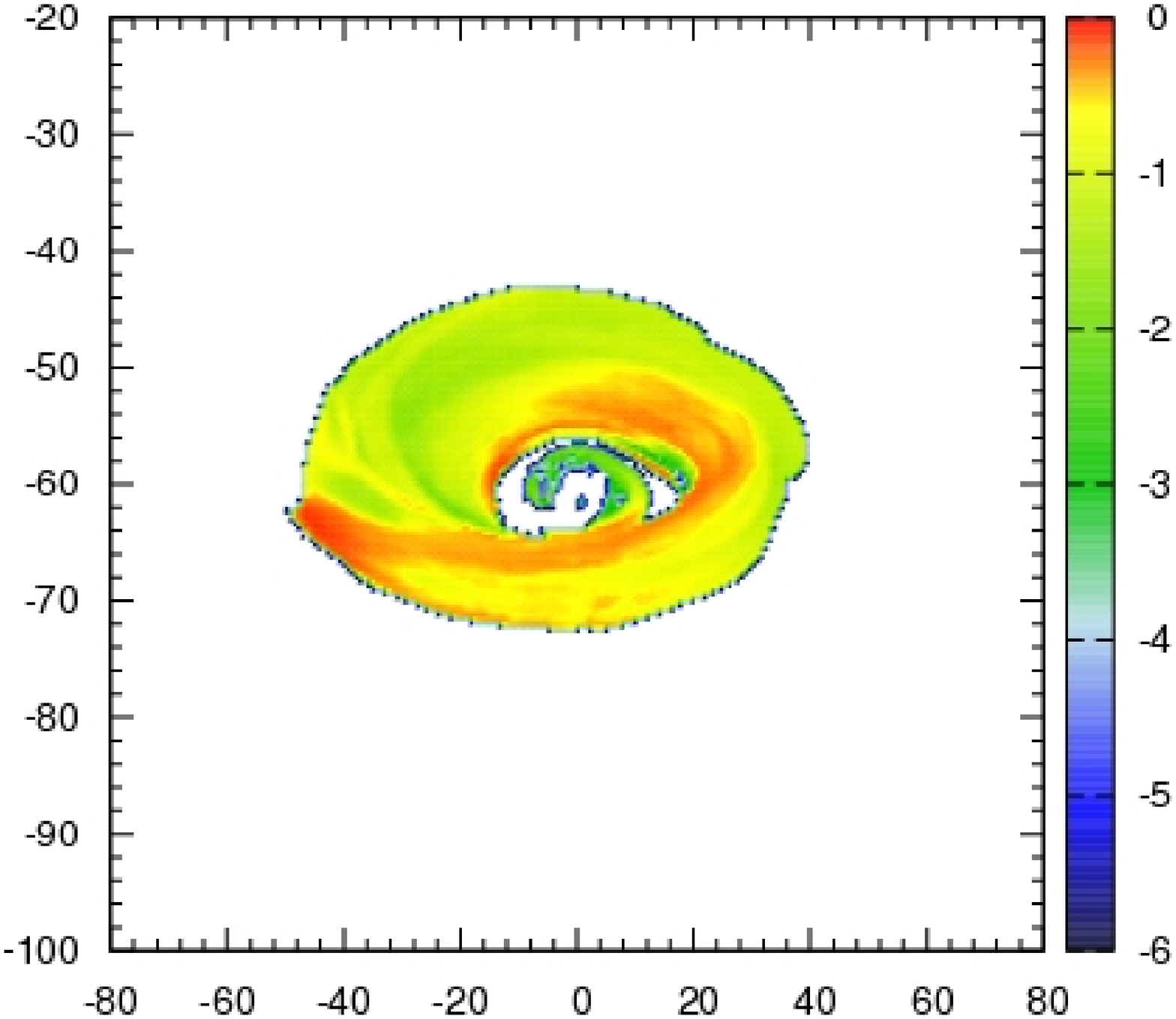}  &
\includegraphics[angle=270,width=2.2in]{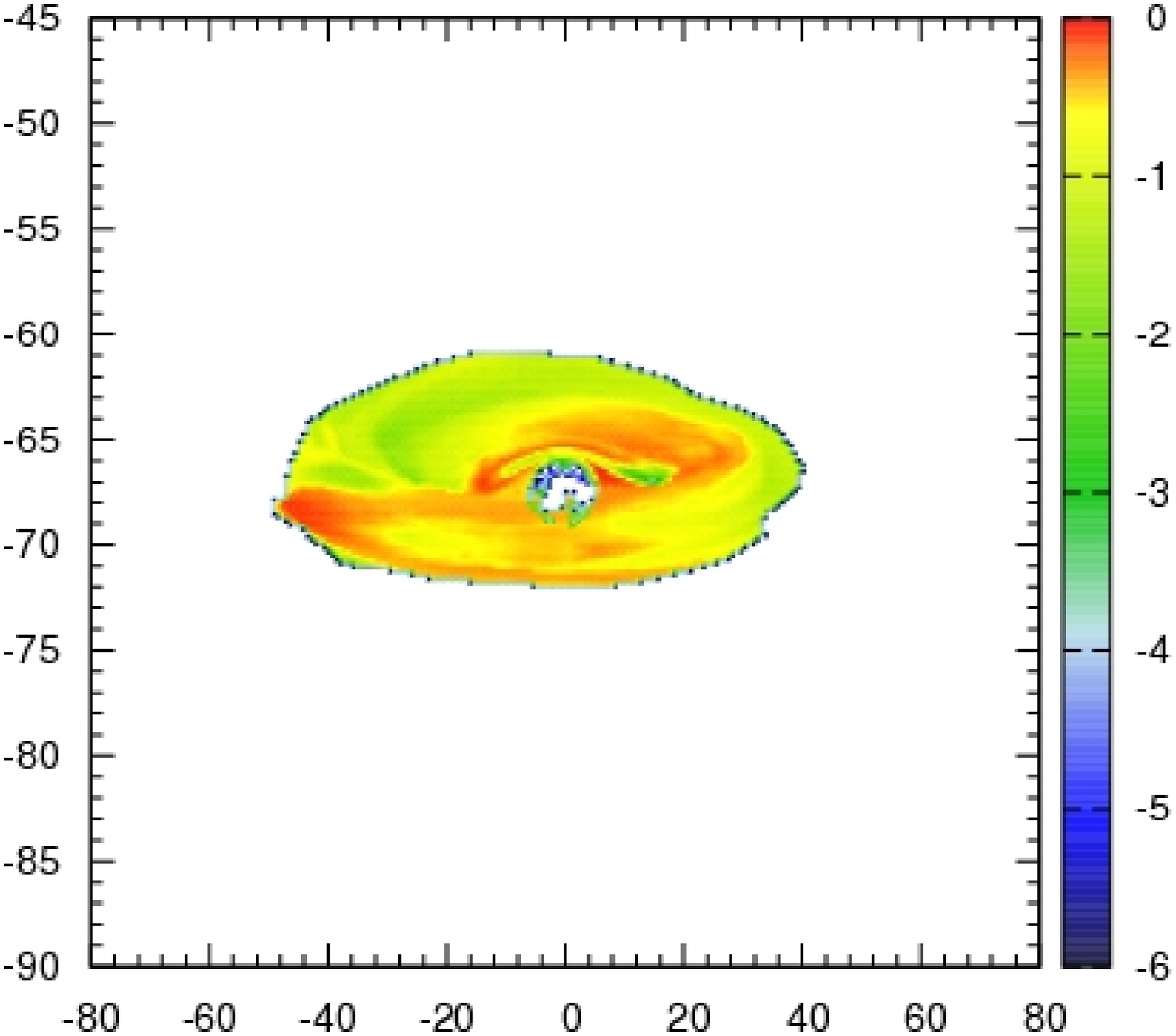}  \\
\includegraphics[angle=270,width=2.2in]{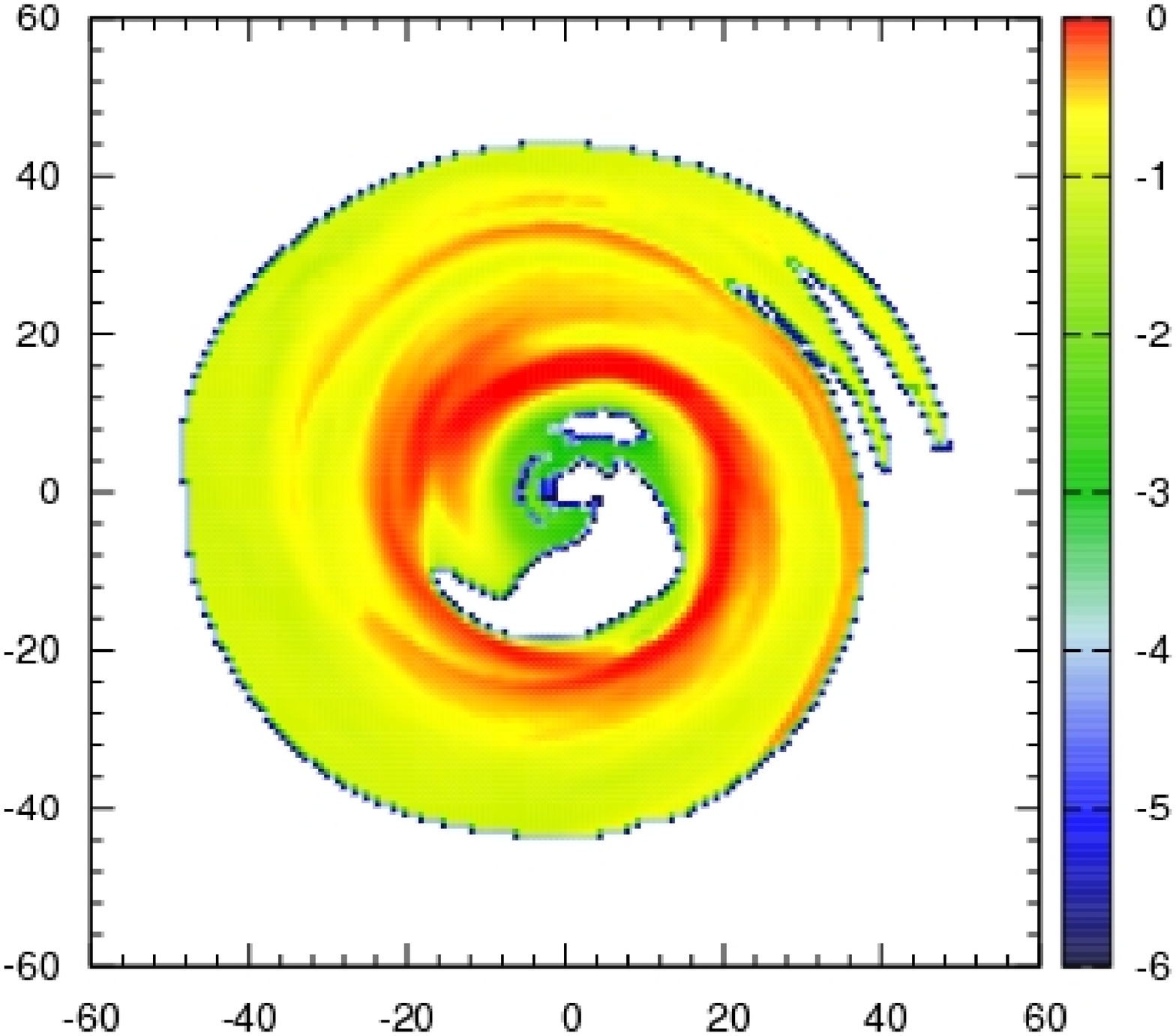}  &  
\includegraphics[angle=270,width=2.2in]{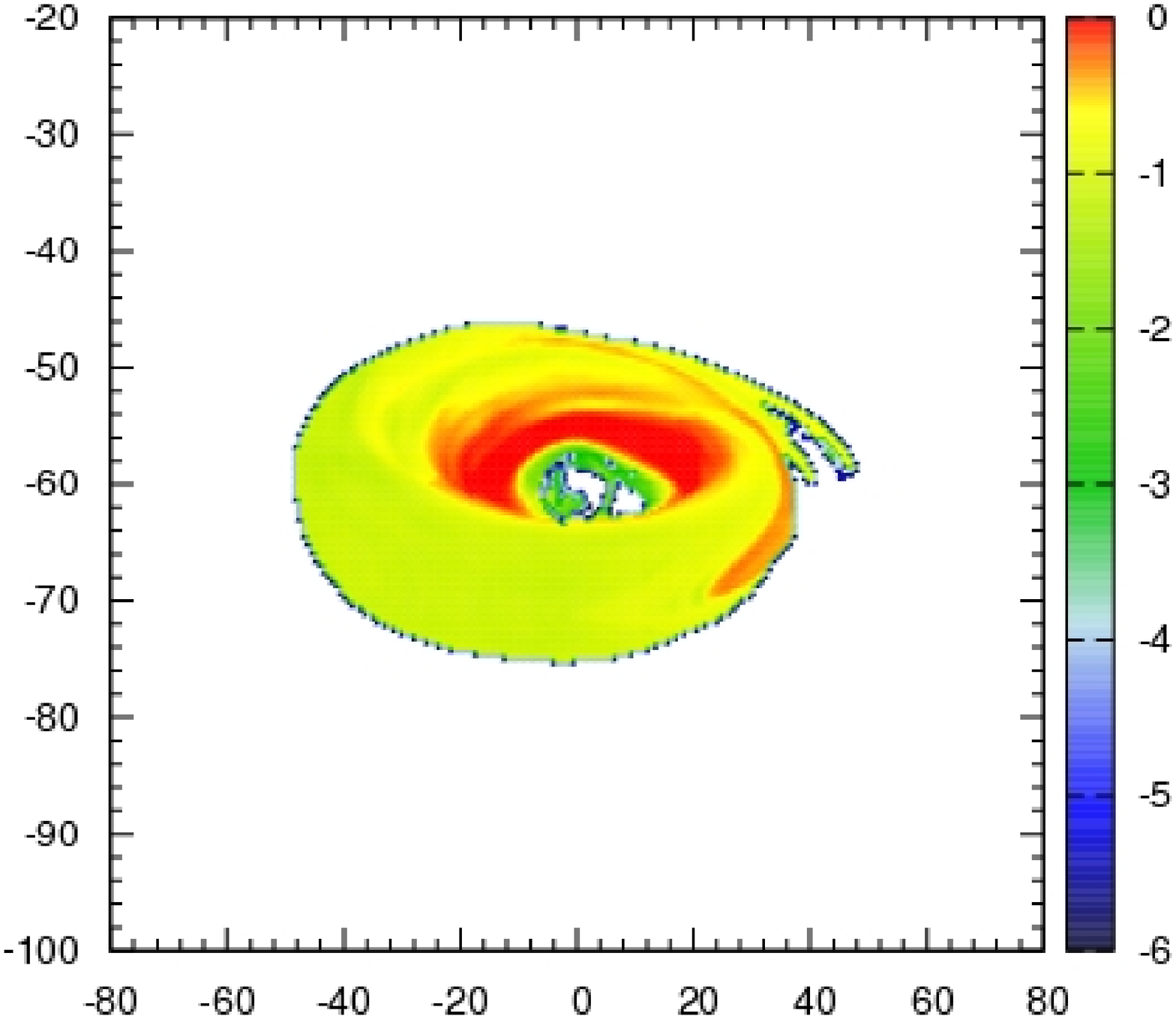}  &
\includegraphics[angle=270,width=2.2in]{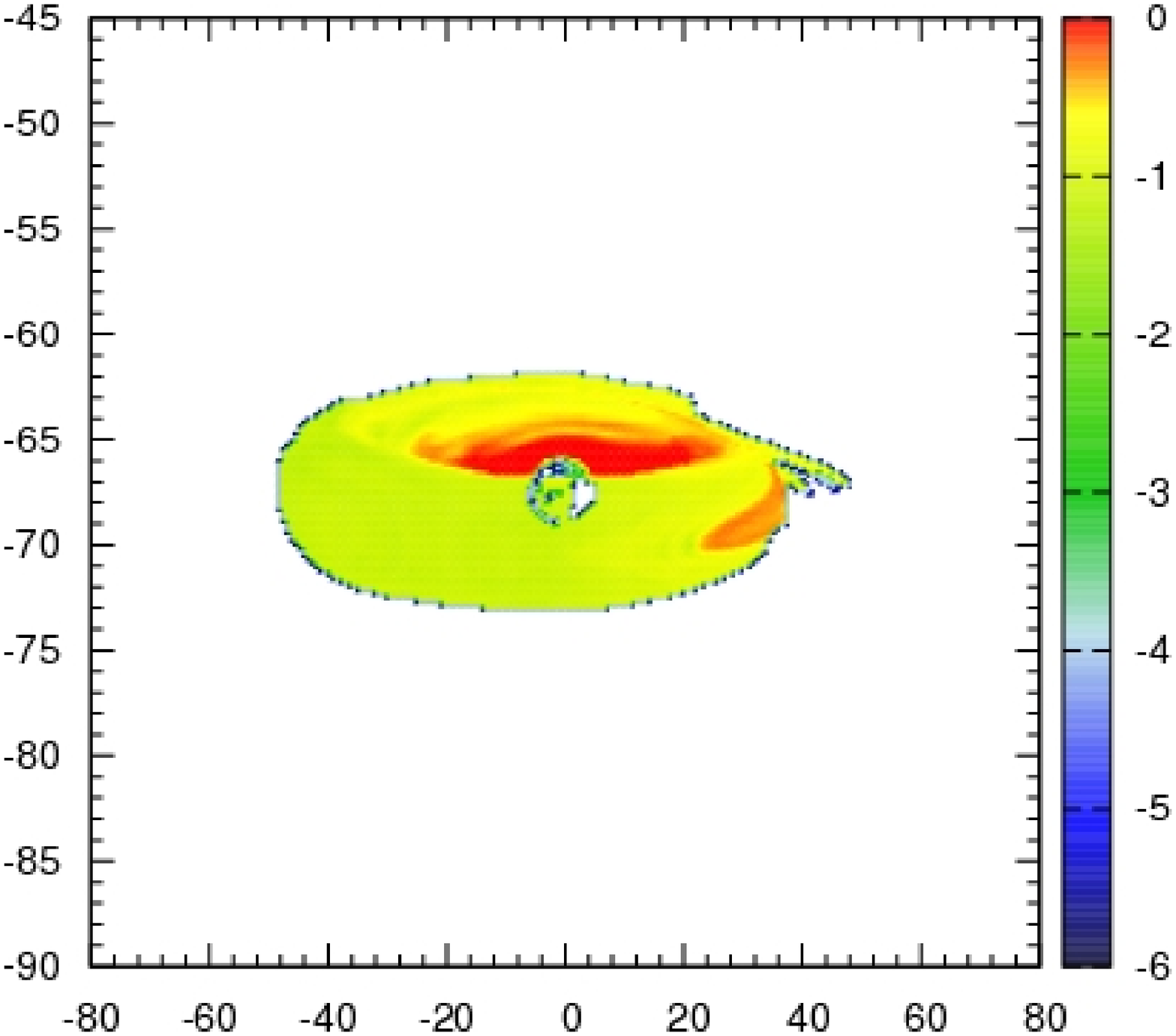}  \\
\includegraphics[angle=270,width=2.2in]{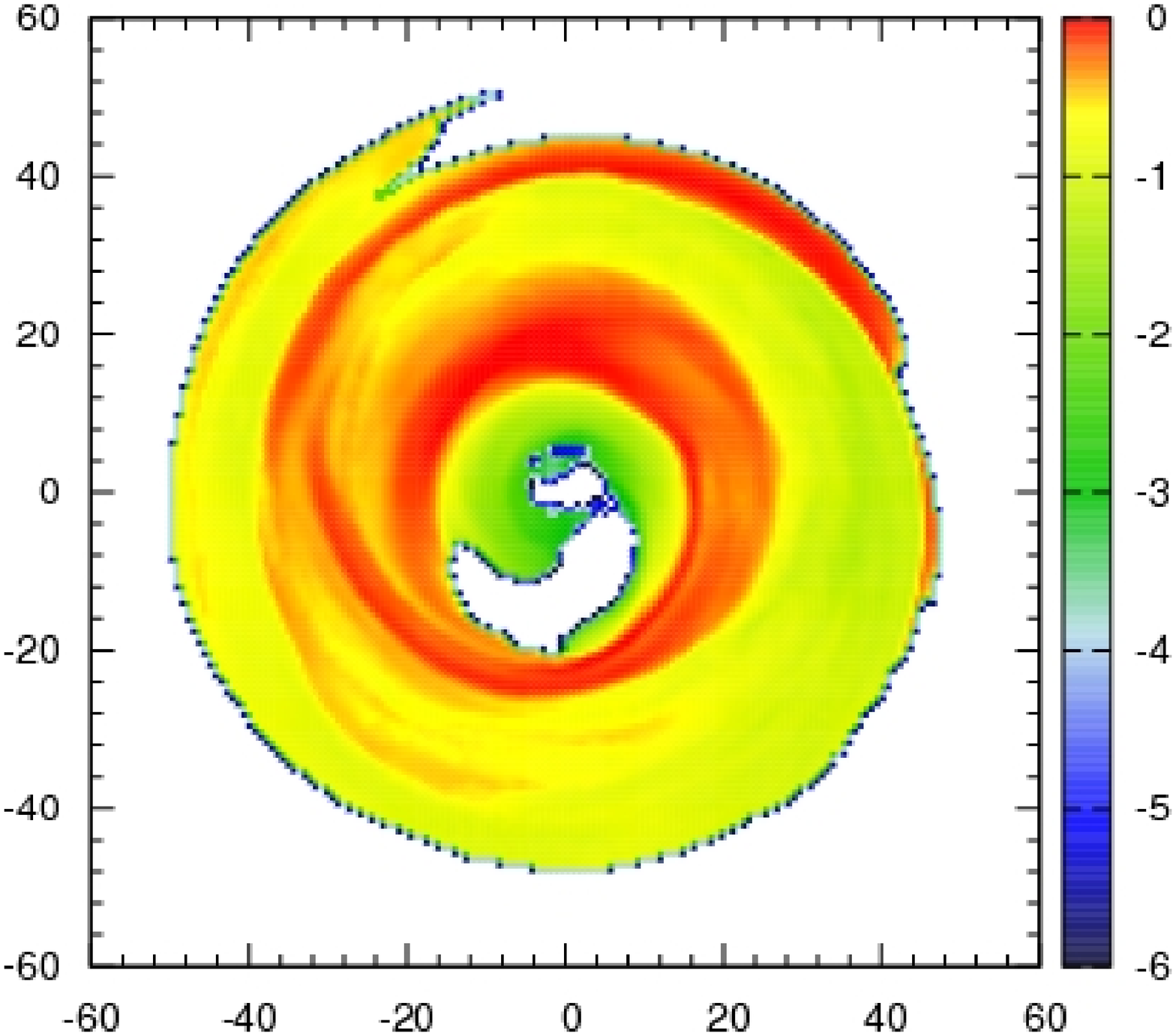}  &  
\includegraphics[angle=270,width=2.2in]{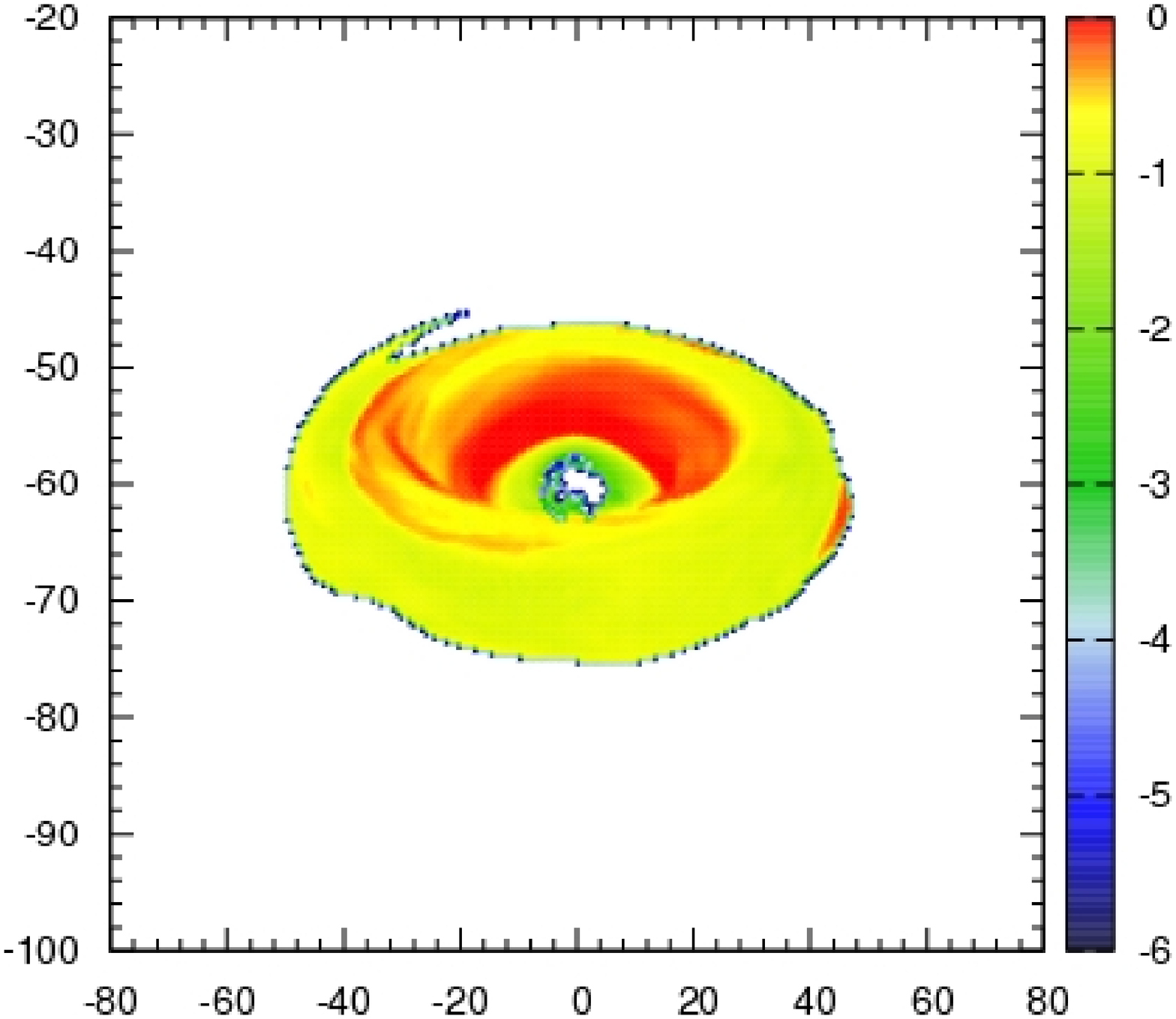}  &
\includegraphics[angle=270,width=2.2in]{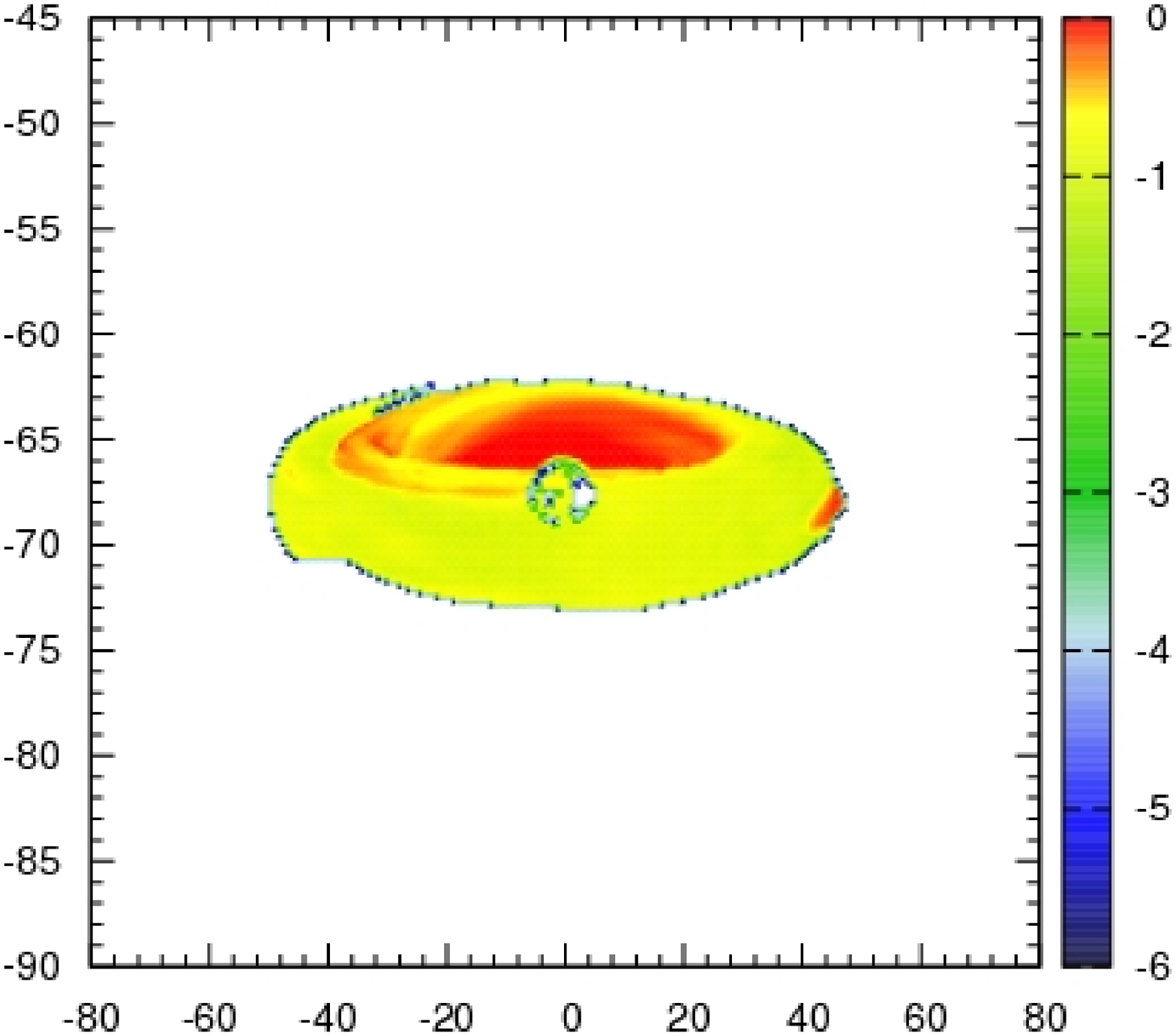}  \\
\end{array}$
\end{center}
\caption{The intensity for $10^{12}$~Hz at $t=1.75 \times 10^5, 1.25\times 10^6, 1.5 \times 10^6,$ and $1.75 \times 10^6$ seconds after the kick
using the ``bremsstrahlung-blackbody'' radiation model. The colormap is a log scale;
The intensity for all the images has been normalized using
the maximum intensity found at $t=1.25\times 10^6$ seconds after the kick.
The images on the left/center/right columns correspond to viewing angles of 
$0^\circ$/$60^\circ$/$75^\circ$ respectively.
The number of geodesics used was 40,000.
The axis scale of the images is in units of mass of the black hole.
\label{f:intensity3}}
\end{figure}
\end{widetext}

%
%
\end{document}